\documentclass[a4paper,11pt]{article}

\usepackage[T1]{fontenc}
\usepackage[latin9]{inputenc}
\usepackage{color}
\usepackage{float}
\usepackage{amsmath,amssymb}
\usepackage{amsthm}
\usepackage{mathrsfs}
\usepackage{graphicx}
\usepackage{bm}
\usepackage{setspace}
\usepackage{esint}
\usepackage{babel}

\newtheorem{theorem}{Theorem}

\newcommand{\mN}{\mathbb{N}}

\newcommand{\Req}[1]{eq.\ (\ref{eq:#1})}

\newcommand{\Leq}[1]{\label{eq:#1}}

\newcommand{\be}{\begin{eqnarray}}
\newcommand{\ee}{\end{eqnarray}}
\newcommand{\ba}{\begin{array}}
\newcommand{\ea}{\end{array}}

\newcommand{\subbe}{\begin{subequations}}
\newcommand{\subee}{\end{subequations}}

\title{Structure Learning in Inverse Ising Problems Using $\ell_2$-Regularized Linear Estimator}

\author{
Xiangming Meng$^{\ddag}$\thanks{Institute for Physics of Intelligence and Department of Physics, Graduate School of Science, The University of Tokyo, 7-3-1, Hongo, Tokyo 113-0033, Japan}
\and 
Tomoyuki Obuchi\thanks{Department of Systems Science, Graduate School of Informatics, Kyoto University, Yoshida Hon-machi, Sakyo-ku, Kyoto-shi, Kyoto 606-8501, Japan}
\and
Yoshiyuki Kabashima$^*$
\thanks{Corresponding author. E-mail: meng@g.ecc.u-tokyo.ac.jp}
}

\begin{document}
\maketitle

\begin{abstract}%
The inference performance of the pseudolikelihood method is discussed in the framework of the inverse Ising problem when the $\ell_2$-regularized (ridge) linear regression is adopted. This setup is introduced for theoretically investigating the situation where the data generation model is different from the inference one, namely the model mismatch situation. In the teacher-student scenario under the assumption that the teacher couplings are sparse, the analysis is conducted using the replica and cavity methods, with a special focus on whether the presence/absence of teacher couplings is correctly inferred or not. The result indicates that despite the model mismatch, one can perfectly identify the network structure using naive linear regression without regularization when the number of spins $N$ is smaller than the dataset size $M$, in the thermodynamic limit $N\to \infty$. Further, to access the underdetermined region $M < N$, we examine the effect of the $\ell_2$ regularization, and find that biases appear in all the coupling estimates, preventing the perfect identification of the network structure. We, however, find that the biases are shown to decay exponentially fast as the distance from the center spin chosen in the pseudolikelihood method grows. Based on this finding, we propose a two-stage estimator: In the first stage, the ridge regression is used and the estimates are pruned by a relatively small threshold; in the second stage the naive linear regression is conducted only on the remaining couplings, and the resultant estimates are again pruned by another relatively large threshold. This estimator with the appropriate regularization coefficient and thresholds is shown to achieve the perfect identification of the network structure even in $0<M/N<1$. Results of extensive numerical experiments support these findings. 
\end{abstract}

%\begin{keywords}
%  inverse Ising problems, structure learning, model mismatch, pseudolikelihood method, replica method
%\end{keywords}

%%%%%%%%%%%%%%%%%%%%%%%%%%%%%%%%%%%%%%%%%%%%%%%%%%%%%
%%%%%%%%%%%%%%%%%%%%%%%%%%%%%%%%%%%%%%%%%%%%%%%%%%%%%
%%%%%%%%%%%%%%%%%%%%%%%%%%%%%%%%%%%%%%%%%%%%%%%%%%%%%
\section{\label{sec:Introduction}Introduction} 
The advent of massive data across various scientific disciplines
has led to widespread use of the classical Ising model as a tool for data modeling \cite{nguyen2017inverse}. Recent applications that have spurred this trend
include retinal neutrons, reconstruction of neural and gene regulatory
networks, and determination of the three-dimensional structure
of proteins in biological sciences \cite{nguyen2017inverse,aurell2012inverse,bachschmid2015learning,berg2017statistical,bachschmid2017statistical,Abbara2019c}.
Inference based on the Ising model is called the inverse Ising problem
or Boltzmann machine learning, which refers to reconstructing the
parameters and structure of an Ising model on the basis of samples of spin configurations.
The maximum likelihood (ML) method is one of the main methods for solving this problem. However, in general, ML is computationally intractable
for a large system. Two popular approaches
have been developed to address this problem. The first is to approximate the ML using approximations
such as Monte Carlo sampling \cite{ackley1985learning,habeck2014bayesian,broderick2007faster} and mean-field approximations \cite{kappen1998efficient,tanaka1998mean,sessak2009small}. The second approach introduces some local
cost function that is easier to optimize instead of
directly maximizing the likelihood function. One of the most effective
examples of the latter is the pseudolikelihood (PL) method \cite{aurell2012inverse,besag1975statistical,decelle2014pseudolikelihood,mozeika2014consistent},
which approximates the likelihood function as the product of conditional
likelihood functions. A prominent advantage of the PL method is that
one can independently estimate the couplings associated with each
spin, as the local couplings directly connected to a single spin
are isolated from the others, thus simplifying the implementation. 

Recently, some theoretical analyses revealing the inference accuracy of the PL method have been conducted using methods of statistical mechanics~\cite{bachschmid2015learning, berg2017statistical,bachschmid2017statistical,Abbara2019c}. For example, in~\cite{bachschmid2017statistical}, assuming that data are drawn independently from an equilibrium Ising model, the learning performance of the PL method with a local cost function was studied for fully connected Ising models using a combination of the replica method and the cavity method~\cite{mezard2009information,opper2001advanced,nishimori2001statistical}.
Subsequently, in~\cite{Abbara2019c}, some of the present authors extended the analysis to sparse couplings. The inverse Ising problem with sparse couplings has a practical relevance in structure learning of graphical models and a number of early studies are found~\cite{aurell2012inverse,decelle2014pseudolikelihood,schmidt2007learning, wainwright2008graphical,ravikumar2010high,santhanam2012information,bresler2015efficiently, vuffray2016interaction,lokhov2018optimal}. These analyses provide a firm theoretical basis for inverse
Ising problems.

In the above studies, the postulated model used in the inference stage covers the true model that generates the data. However, such an assumption does not necessarily hold in practical situations, because the data-generating model is generally unknown \textit{a priori}. Therefore, it is important to evaluate the learning performance of the popular PL method in model mismatch cases, which is the main focus of this study. 

Specifically, within the teacher-student scenario, we examine the inference performance of the PL method when $\ell_2$-regularized (ridge) linear regression is applied to data generated from the teacher Ising model with sparse couplings. The ridge regression is very simple but widely used in practical situations of data analysis, and thus is appropriate as a starting point for the present purpose. Our main question is whether the presence/absence of teacher coupling can be correctly inferred or not even in this mismatched case. To answer this question, we employ similar analytical techniques to those in~\cite{Abbara2019c}: we use the replica and cavity methods and take the thermodynamic limit where the number of spins $N$ goes to infinity and the dataset size $M$ is proportional to $N$ as $M=\alpha N$ with $\alpha=\mathcal{O}(1)$; furthermore we assume the tree-like structure of the network of the teacher couplings and generalize the ansatz in~\cite{Abbara2019c} about the mean estimates of couplings on an assumed support. This generalized ansatz enables us to systematically treat the effect of the regularization. 

As a result, we find that the $\ell_2$ regularization causes undesirable biases in the overall coupling estimates while without regularization no such biases exist for the estimates on the set of absent teacher couplings. This indicates that for $\alpha=M/N > 1$ the perfect recovery of the network structure is possible by the naive linear regression without regularization, since the fluctuation of the estimates vanishes in the thermodynamic limit as in the matched case~\cite{Abbara2019c} and thus we can effectively prune false positive couplings by a reasonable threshold. Meanwhile in the case of $\alpha \leq 1$, the regularization is necessary for obtaining the estimates and thus the biases are unavoidable, which makes the perfect recovery difficult. To overcome this, we further quantitatively analyze those biases, and find that they decay exponentially fast as the distance from the center spin chosen in the PL method grows. This finding motivates us to introduce a two-stage estimator for systematically achieving the perfect recovery even for $\alpha\leq 1$. The actual procedures of this two-stage estimator are as follows. In the first stage, we perform the ridge regression and then prune the estimates with a certain threshold to obtain $\mathcal{O}(1)$ coupling estimates\footnote{To facilitate perfect recovery, this threshold should be sufficiently small to ensure full recall. If this threshold is independent of the system size $N$, then the number of false positive couplings is expected to be $\mathcal{O}(1)$ if $N$ is large enough, because the majority of the spins is far from the center spin in the PL method if we assume the tree-like network and thus the corresponding biases are negligibly small compared to the threshold.}. In the second stage, the naive linear regression without
regularization is performed only on the remaining couplings at the end of the first stage: the number of remaining couplings is expected to be $\mathcal{O}(1)$ and thus the dataset size is effectively very large ($\alpha\gg 1$) and hence we can again eliminate false positive couplings by a reasonable threshold. These results thus finally provide a positive answer to our question: the perfect recovery is possible as long as $\alpha>0$ even in the mismatched case! To support these analytical results, we also conduct numerical experiments on the random regular (RR) graph and the Erd\H{o}s--R\'enyi (ER) graph. The result is fairly consistent with all the analytical predictions and thus supports our findings. 

The remainder of this paper is organized as follows. Section \ref{sec:Inverse-Ising-problem} reviews the inverse Ising problem and some typical estimators. In addition, it presents the problem setup in the teacher-student scenario. Section \ref{sec:Statistical-mechanical-analysis} describes the statistical mechanics analysis of the $\ell_{2}$-regularized linear estimator, drawing on previous studies \cite{bachschmid2017statistical,Abbara2019c} for sparse couplings. Numerical simulations are conducted to evaluate the accuracy of the theoretical analysis, and Section \ref{sec:Numerical-experiments} compares the experimental results with the theoretical analysis. Finally, Section \ref{sec:Summary-and-discussion} concludes the paper. 

%%%%%%%%%%%%%%%%%%%%%%%%%%%%%%%%%%%%%%%%%%%%%%%%%%%%%
%%%%%%%%%%%%%%%%%%%%%%%%%%%%%%%%%%%%%%%%%%%%%%%%%%%%%
%%%%%%%%%%%%%%%%%%%%%%%%%%%%%%%%%%%%%%%%%%%%%%%%%%%%%
\section{\label{sec:Inverse-Ising-problem}Inverse Ising Problem}
Let us consider an Ising model with $N$ binary spin variables $\boldsymbol{s}=\left( s_{i}=\pm1\right)_{i=0}^{N-1}$, which follows the Boltzmann distribution
\begin{equation}
P_{\textrm{Ising}}\left(\boldsymbol{s}|\boldsymbol{J,H}\right)=\frac{1}{Z_{\textrm{Ising}}}e^{\sum_{i<j}J_{ij}s_{i}s_{j}+\sum_{i}H_{i}s_{i}},\label{eq:BotzmanDistr-def}
\end{equation}
where $Z_{\textrm{Ising}}$ is the partition function and $\boldsymbol{J}=\left( J_{ij}\right)_{ij} \in\mathbb{R}^{N\times N}$ and $\boldsymbol{H}=\left( H_{i}\right)_{i=0}^{N-1} \in\mathbb{R}^{N}$ are the couplings and external fields, respectively. In (\ref{eq:BotzmanDistr-def}), the temperature is absorbed in $\boldsymbol{J}$ and $\boldsymbol{H}$. The standard goal of the inverse Ising problem is to learn the couplings $\boldsymbol{J}$ and external fields $\boldsymbol{H}$ from a set of observations of spin snapshots $\mathcal{D}^{M}=\left\{ \boldsymbol{s}^{\left(\mu\right)}\right\} _{\mu=1}^{M}$, where $M$ denotes the number of samples in the dataset, i.e., dataset size. Especially, a particular interest is on learning the network structure composed of the couplings. Our main focus in this paper is to reveal whether the structure learning is possible or not based on the linear estimator with the $\ell_2$ regularization, as detailed below.  

%%%%%%%%%%%%%%%%%%%%%%%%%%%%%%%%%%%%%%%%%%%%%%%%%%%%%
%%%%%%%%%%%%%%%%%%%%%%%%%%%%%%%%%%%%%%%%%%%%%%%%%%%%%
\subsection{\label{sec:Some-Estimators}Some Estimators}
Here we summarize some estimators for the inverse Ising problem and also describe the motivations for evaluation of the linear estimator. 

%%%%%%%%%%%%%%%%%%%%%%%%%%%%%%%%%%%%%%%%%%%%%%%%%%%%%
\subsubsection{\label{sec:Maximum-Likelihood-Estimator}Maximum Likelihood Estimator}
The canonical estimator in statistics is the one based on the maximum likelihood (ML) method and is defined as
\begin{equation}
\left\{ \hat{\boldsymbol{J}}^{ML},\hat{\boldsymbol{H}}^{ML}\right\} =\underset{\boldsymbol{J,H}}{\arg\min}\left\{ -\sum_{\mu=1}^{M}\log P_{\textrm{Ising}}\left(\boldsymbol{s}^{\left(\mu\right)}|\boldsymbol{J,H}\right)\right\}.\label{eq:ML-def}
\end{equation}
This shows some useful properties such as consistency and asymptotic efficiency. However in the inverse Ising problem, the ML method suffers from the high computational complexity because the exponentially large computational cost with respect to (w.r.t.) $N$ is needed to compute $Z_{\rm Ising}$. Due to this limitation, other estimators than the ML one are usually practically chosen.

%%%%%%%%%%%%%%%%%%%%%%%%%%%%%%%%%%%%%%%%%%%%%%%%%%%%%
\subsubsection{\label{sec:Maximum-Pseudolikelihood-Estimator}Maximum Pseudolikelihood Estimator}
An alternative to the ML method is the pseudolikelihood (PL) method \cite{besag1975statistical}, which replaces the original likelihood with the conditional distribution $P\left(s_{i}|\boldsymbol{s}_{\setminus i},\boldsymbol{J}_{i},H_{i}\right)$ for each spin $s_{i}$, where $\boldsymbol{J}_{i}=\left( J_{ij}\right) _{j(\neq i)}$ is the coupling vector connected to spin $s_{i}$ and $\boldsymbol{s}_{\setminus i}$ is the spin vector $\boldsymbol{s}$ excluding $s_{i}$. Specifically, for each $i$, the conditional distribution $P\left(s_{i}|\boldsymbol{s}_{\setminus i},\boldsymbol{J}_{i},H_{i}\right)$ is of the form
\begin{equation}
P\left(s_{i}|\boldsymbol{s}_{\setminus i},\boldsymbol{J}_{i},H_{i}\right)=\frac{1}{Z_{i}}e^{s_{i}\left(\sum_{j(\neq i)}J_{ij}s_{j}+H_{i}\right)},\label{eq:PL-def}
\end{equation}
where $Z_{i}=2\cosh\left(\sum_{j(\neq i)}J_{ij}s_{j}+H_{i}\right)$ is the site partition function. Consequently, the PL estimator is applied to each $i$ separately, leading to
\begin{align}
\left\{ \hat{\boldsymbol{J}_{i}}^{PL},\hat{H_{i}}^{PL}\right\}  =\underset{\boldsymbol{J}_{i},H_{i}}{\arg\min}\left\{ \sum_{\mu=1}^{M} \left(- s_{i}^{\left(\mu\right)}h_{i}\left(\boldsymbol{s}_{\setminus i}^{\left(\mu\right)},\boldsymbol{J}_{i},H_{i}\right) + \log2\cosh(h_{i}\left(\boldsymbol{s}_{\setminus i}^{\left(\mu\right)},\boldsymbol{J}_{i},H_{i}\right))\right) \right\},\label{eq:PL-estimator-df}
\end{align}
where $h_{i}\left(\boldsymbol{s}_{\setminus i}^{\left(\mu\right)},\boldsymbol{J}_{i},H_{i}\right) =\sum_{j(\neq i)}J_{ij}s_{j}+H_{i}$. 

The PL method has two remarkable properties: consistency and locality \cite{hyvarinen2006consistency}. Consistency means that the PL estimator converges to the true value when the dataset size $M$ is sufficiently large. Locality means that each coupling vector $\boldsymbol{J}_{i}$ can be estimated independently, which leads to low computational complexity. For obtaining the coupling estimates for all couplings, the PL estimator should be computed for all $i=1,\cdots,N$ separately.

According to earlier studies~\cite{Abbara2019c,schmidt2007learning,wainwright2008graphical,ravikumar2010high}, the perfect recovery of the network structure is possible by this PL estimator. Its information theoretic limit when employed with the $\ell_1$ regularization is derived in~\cite{ravikumar2010high}, showing that the perfect recovery is possible in the large $N$ limit satisfying $M > k \log N$ with an appropriate constant $k>0$. Meanwhile, when the regularization is absent, the perfect recovery is again shown to be possible in the large $N$ limit satisfying $\alpha=M/N>2$ in~\cite{Abbara2019c}. In the latter study, the direct values of the variance and bias of the estimator are computed by using the statistical mechanical methods, and we employ the same approach for analyzing the performance of the linear estimator in this study. 

%%%%%%%%%%%%%%%%%%%%%%%%%%%%%%%%%%%%%%%%%%%%%%%%%%%%%
\subsubsection{\label{sec:Linear-Estimator}Linear Estimator}
The simplest estimator in regression is linear one. Thus we propose a linear estimator for the inverse Ising problem as follows:
\begin{align}
\{ \hat{\boldsymbol{J}_{i}} ,\hat{H}_i \}
& 
=
\underset{\boldsymbol{J}_{i},H_i}{\arg\min}
\left\{ 
\sum_{\mu=1}^{M}
\left( s_{i}^{\left(\mu\right)}-\sum_{j(\neq i)}J_{ij}s_{j}^{\left(\mu\right)} -H_i \right)^{2}
+\lambda N\sum_{j(\neq i)}J_{ij}^{2}
\right\}. \label{eq:PL-estimator-L2}
% & =\underset{\boldsymbol{J}_{i}}{\arg\min}\left\{ \left\Vert \boldsymbol{s}_{i}-\boldsymbol{A}\boldsymbol{J}_{i} 
% - H_i \bm{1}_M
% \right\Vert _{2}^{2}+\lambda N\left\Vert \boldsymbol{J}_{i}\right\Vert _{2}^{2}\right\} ,
\end{align}
%%
% where we introduced the notations $\boldsymbol{s}_{i}=\left( s_{i}^{\left(\mu\right)}\right) _{\mu=1}^{M}\in\mathbb{R}^{M\times1}$, $\boldsymbol{A}=\left( s_{j}^{\left(\mu\right)}\right) _{\mu,j(\neq i)}\in\mathbb{R}^{M\times\left(N-1\right)}$, and $\bm{1}_M=(1,\cdots,1)^{\top}\in \mR^{M \times 1}$ for later notational convenience. 
As for the PL method, we focus on a single spin and perform the learning locally also in this case. The $\ell_2$ regularization is introduced to make the estimator well defined even in the underdetermined situation $\alpha<1$ and the factor $N$ is introduced for the scaling to be appropriate.

This estimator implies that the corresponding inference model is outside the parameter family of the generative model, and hence the model mismatch occurs. Consequently, this estimator does not show consistency. It thus becomes more nontrivial whether the perfect recovery of the network structure is possible or not. Since the linear estimator is largely superior to the ML and PL ones in terms of the computational complexity/analytical amenability, its advantage will be huge if the perfect recovery is shown to be possible even by this linear estimator. Below we tackle this problem, to eventually find a positive answer. 

%%%%%%%%%%%%%%%%%%%%%%%%%%%%%%%%%%%%%%%%%%%%%%%%%%%%%
%%%%%%%%%%%%%%%%%%%%%%%%%%%%%%%%%%%%%%%%%%%%%%%%%%%%%
\subsection{Problem Setup: Linear Estimator in Teacher-Student Scenario}
In this paper, we investigate the properties of the above linear estimator in the teacher-student scenario. The dataset $\mathcal{D}^{M}=\left\{ \boldsymbol{s}^{\left(\mu\right)}\right\} _{\mu=1}^{M}$ is assumed to be generated independently from a teacher Ising model with couplings $\boldsymbol{J}^{*}$ and external fields $\boldsymbol{H}^{*}$. We denote by $\left[\cdot\right]_{\mathcal{D}^{M}}$ the expectation over the dataset $\mathcal{D}^{M}$ generated in this way, i.e.,
\begin{equation}
\left[\cdot\right]_{\mathcal{D}^{M}}=\sum_{\boldsymbol{s}^{\left(1\right)},...,\boldsymbol{s}^{\left(M\right)}}\left(\cdot\right)\prod_{\mu=1}^{M}P_{\textrm{Ising}}\left(\boldsymbol{s}^{\left(\mu\right)}|\boldsymbol{J}^{*},\boldsymbol{H}^{*}\right).\label{eq:quenched-average-def}
\end{equation}
For simplicity of analysis, the external fields are assumed to be zero in the following, i.e., $\boldsymbol{H}^{*}=0$. Furthermore, we assume the teacher couplings' network is tree-like as in \cite{Abbara2019c}. Representative examples of such networks are the RR graph and the ER graph with small edge probability. Our main focus is on whether we can recover this network structure based on the linear estimator, or not. 

Correspondingly, we mainly analyze the following three quantities related to the structure learning: the residual sum of square (RSS)
\begin{equation}
\mathcal{E}
=
\left\Vert \boldsymbol{J}_{i}^{*}-\hat{\boldsymbol{J}_i}\right\Vert _{2}^{2},
%=\left[\left\Vert \boldsymbol{J}_{i}^{*}-\hat{\boldsymbol{J}_i}\right\Vert _{2}^{2}\right]_{\mathcal{\mathcal{D}^{M}}},
\label{eq:RSS-def-1}
\end{equation}
the variance of the estimator, and the rates of correctly inferring the presence/absence of couplings. For judging the presence/absence of couplings, a judging scheme is needed and we implement this by thresholding the estimator: we introduce a certain threshold $K_{\rm th}$ and if $|\hat{J}_i|>K_{\rm th}$ then we judge the corresponding coupling is present, otherwise it is supposed to be absent. The true positive rate, the rate of correctly inferred to be present among the present couplings, is denoted as $TP$. Similarly, the true negative, false positive, and false negative rates are denoted by $TN$, $FP$, and $FN$ respectively. {\it Precision} and {\it Recall}, common statistical measures of inference accuracy, are defined by these quantities as
\be
{\rm Precision}=\frac{TP}{TP+FP},~{\rm Recall}=\frac{TP}{TP+FN},
\Leq{Precision-Recall}
\ee
and we quantify the network recovery accuracy by these two quantities. The reason why we do not directly use $TP$ and $FP$ is the imbalance in the presence rate of couplings since we assume the sparse network. 

Below we state how these quantities are computed by the statistical mechanical analysis.

%%%%%%%%%%%%%%%%%%%%%%%%%%%%%%%%%%%%%%%%%%%%%%%%%%%%%
%%%%%%%%%%%%%%%%%%%%%%%%%%%%%%%%%%%%%%%%%%%%%%%%%%%%%
%%%%%%%%%%%%%%%%%%%%%%%%%%%%%%%%%%%%%%%%%%%%%%%%%%%%%
\section{\label{sec:Statistical-mechanical-analysis}Statistical Mechanical Analysis}
In this section, we present the statistical mechanical analysis of the inference performance of the $\ell_2$-regularized linear estimator (\ref{eq:PL-estimator-L2}) following the previous studies \cite{bachschmid2017statistical,Abbara2019c}. In the following, we first present the statistical mechanical formulation of the problem, illustrating its basic idea and difficulty. Afterwards, details of how to tackle such difficulty are illustrated in Sections 3.1-3.4. 

For simplicity and without loss of generality, we denote the index of the focused spin as $0$ and the coupling vector to be inferred by $\boldsymbol{J}$, where the index $0$ is omitted.  Following the standard prescription of statistical mechanics, the Hamiltonian corresponding to the cost function (\ref{eq:PL-estimator-L2}) is 
\begin{align}
\mathcal{H}\left(\boldsymbol{J}|\mathcal{D}^{M}\right) & =\sum_{\mu=1}^{M}\left(s_{0}^{\left(\mu\right)}-\sum_{j=1}^{N-1}J_{j}s_{j}^{\left(\mu\right)}\right)^{2}+\lambda N\sum_{j=1}^{N-1}J_{j}^{2}\nonumber \\
 & =\sum_{\mu=1}^{M}\Phi\left(s_{0}^{\left(\mu\right)}h^{\left(\mu\right)}\right)+\lambda N\sum_{j=1}^{N-1}J_{j}^{2},\label{eq:Hamilton}
\end{align}
where $\Phi\left(x\right)=\left(x-1\right)^{2}$ and $h^{\left(\mu\right)}=\sum_{j=1}^{N-1}J_{j}s_{j}^{\left(\mu\right)}$. Then, the Gibbs-Boltzmann distribution of the student couplings is defined as
\begin{equation}
P\left(\boldsymbol{J}|\mathcal{D}^{M}\right)=\frac{1}{Z}\exp\left[-\beta\mathcal{H}\left(\boldsymbol{J}|\mathcal{D}^{M}\right)\right],\label{eq:Gibbs-distribution}
\end{equation}
where $\beta$ represents the inverse temperature and $Z$ is the partition function
\begin{equation}
Z=\int d\boldsymbol{J}\exp\left[-\beta\mathcal{H}\left(\boldsymbol{J}|\mathcal{D}^{M}\right)\right].\label{eq:partition Z}
\end{equation}
The Gibbs-Boltzmann distribution becomes the point-wise measure on the solution of (\ref{eq:PL-estimator-L2}) in the zero-temperature limit $\beta\rightarrow+\infty$, meaning that we can extract any information of the estimator from the Gibbs-Boltzmann distribution or the free energy. Hence, we concentrate on computing the free energy in the zero-temperature limit in the following. The free energy density averaged over the dataset is given by
\begin{equation}
f=-\frac{1}{N\beta}\left[\log Z\right]_{\mathcal{D}^{M}}. \label{eq:free energy def}
\end{equation}
Unfortunately, the average over the dataset of $\log Z$ is analytically difficult. To overcome this, we use the replica method from the statistical mechanics of disordered systems~\cite{mezard2009information,opper2001advanced,nishimori2001statistical} as
\begin{align}
f & =-\frac{1}{N\beta}\left[\log Z\right]_{\mathcal{D}^{M}}=-\lim_{n\rightarrow0}\frac{1}{N\beta}\frac{\partial}{\partial n}\log\left[Z^{n}\right]_{\mathcal{D}^{M}},\label{eq:F-replica-def}\\
\left[Z^{n}\right]_{\mathcal{D}^{M}} & =\int\prod_{a=1}^{n}d\boldsymbol{J}^{a}e^{-\beta \lambda N\sum_{a=1}^{n}\left\Vert \boldsymbol{J}^{a}\right\Vert _{2}^{2}}\left\{ \sum_{\boldsymbol{s}}P_{\textrm{Ising}}\left(\boldsymbol{s}|\boldsymbol{J}^{*}\right)\exp\left[-\beta\sum_{a=1}^{n}\Phi\left(s_{0}h^{a}\right)\right]\right\} ^{\alpha N}, \label{eq:Zn-replicated partition}
\end{align}
where $P_{\textrm{Ising}}\left(\boldsymbol{s}|\boldsymbol{J}^{*}\right)$ is the Boltzmann distribution of the teacher network in (\ref{eq:BotzmanDistr-def}) with $\boldsymbol{H}^{*}=0$ and the so-called cavity field is introduced:
\be
h^a=\sum_{j=1}^{N-1}J_{j}^{a}s_j.
\ee
According to the standard prescription of the replica method, in \Req{Zn-replicated partition} we assumed $n\in \mN$ to proceed with the calculation. The limit $n\to 0$ in \Req{F-replica-def} is taken by using an analytical continuation of this expression at the end. To find such an expression is the task below.

%%%%%%%%%%%%%%%%%%%%%%%%%%%%%%%%%%%%%%%%%%%%%%%%%%%%%
%%%%%%%%%%%%%%%%%%%%%%%%%%%%%%%%%%%%%%%%%%%%%%%%%%%%%
\subsection{\label{subsec:Sparse-Anatz}Ansatz For Handling Cavity Fields } 
To calculate the integration in (\ref{eq:Zn-replicated partition}), we resort to the cavity approach used in \cite{bachschmid2017statistical,Abbara2019c}. As the case of~\cite{Abbara2019c}, the cavity field $h^{a}=\sum_{j}J_{j}^{a}s_{j}$ obeys a nontrivial distribution in the present case. To address this problem, we propose an ansatz which is a generalization of the one used in \cite{Abbara2019c}. The generalized ansatz enables us to systematically treat the estimation bias on the coupling estimates in the inactive set $ \{i|J_i^*=0,i\in\{1,\cdots,N-1\} \}$. Such biases are absent in~\cite{Abbara2019c} but present in our case due to the $\ell_2$ regularization we employ. In this subsection the details of the generalized ansatz are explained.

\begin{figure}[thb]
\begin{center}
\includegraphics[scale=0.35]{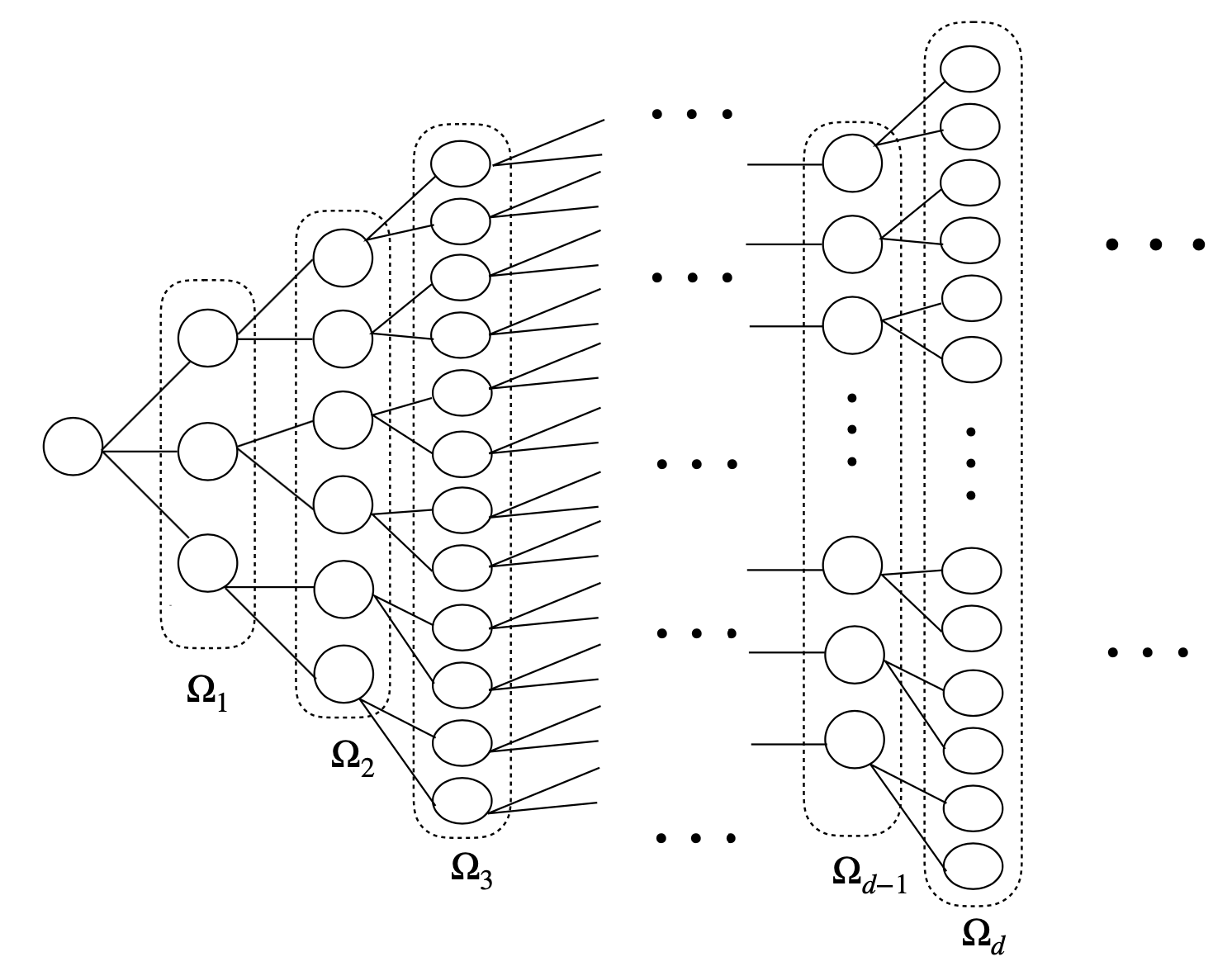}
\caption{Schematic of generations of spins. In general, the $d$-th generation of spin $s_{0}$ is denoted as $\Omega_{d}$, whose distance from spin $s_{0}$ is $d$. \label{fig:tree-graph-NN}}
\end{center}
\end{figure}

The basic idea of the ansatz is to categorize the estimators based on the distance or generation from the focused spin $s_0$. If we consider a teacher Ising model whose coupling network takes a tree-like graph, we can naturally define generations of the spins according to the distance from the focused spin $s_0$. We categorize the spins directly connected to $s_0$ as the first generation and denote the corresponding index set as $\Omega_{1}=\{ i|J_{i}^{*}\neq0,i\in  \left\{ 1,\ldots,N-1\right\} \} $. Each spin in $\Omega_1$ is connected to some other spins except for $s_0$, and those spins constitute the second generation and we denote its index set as $\Omega_2$. This recursive construction of generations can be unambiguously continued on the tree-like graph, and we denote the index set of the $d$-th generation from spin $s_{0}$ as $\Omega_{d}$. The overall construction of generations is graphically represented in Fig. \ref{fig:tree-graph-NN}. 

Let us state our ansatz using the above definitions and notations. We introduce $\Psi_d=\left\{ \Omega_{1},\Omega_{2}\ldots,\Omega_{d}\right\} $ and call it the nearest neighbors (NN) set of $d$ generations. Our ansatz assumes that the estimates $\boldsymbol{\hat{J}}=\left(\hat{J_{i}}\right)_{i=1}^{N-1}$ obey the following form: 
\begin{equation}
\hat{J_{i}}=\begin{cases}
\text{\ensuremath{\bar{J}_{i}}}+\frac{1}{\sqrt{N}}\Delta_{i}, & i\in\Psi_d,\\
\frac{1}{\sqrt{N}}\Delta_{i}, & i\in\bar{\Psi}_d,
\end{cases}\label{eq:W-approximate}
\end{equation}
where $\bar{\Psi}_d$ denotes the complement set of $\Psi_d$ and $\Delta_{i}$ is a random variable whose mean and variance are zero and $\mathcal{O}(1)$, respectively. The variance of the estimator corresponds to the variance of $\Delta_i/\sqrt{N}$, and thus it shrinks in the scaling $\mathcal{O}(1/N)$ in the large $N$ limit. This is important for structure learning as later mentioned in Section 4. We call $\{ \bar{J}_i \}_{i\in \Psi_d}$ mean estimates which are determined by minimizing the free energy. In this sense, the mean estimates can be considered as order parameters. Meanwhile, $\{ \Delta_{i} \}_{i}$ are termed noise variables and also are integration variables which replace the ones in (\ref{eq:Zn-replicated partition}). If the estimation bias is absent in the inactive set $\bar{\Omega}_1$, then $d=1$ is sufficient to take into account all the non-zero means in the estimators and thus is exact. This is the case in the earlier study~\cite{Abbara2019c}. In the present case, however, the estimation bias exists also in the inactive set and hence we need larger values of $d$. In general, we can expect that the approximation of the free energy will be more accurate as $d$ grows, but it involves the exponential increase of the number of the order parameter $\{\bar{J}_i\}_{i\in \Psi_d}$. Fortunately, as we see later, some small values of $d$, say $d=2$, provide a large improvement from $d=1$ and a quantitatively satisfactory result. This good nature comes from the fact that the absolute values of the mean estimates $\{\bar{J}_i\}_{i \in \Psi_d}$ decay exponentially fast as $d$ increases, which is proved in Section \ref{subsec:Nearest-Neighours-Effect} in the framework of the replica method. In this way we can provide an accurate ansatz to handle the integration in (\ref{eq:Zn-replicated partition}). The calculation details will be shown in the next subsections, and in the remaining part of this subsection we discuss some consequences of the ansatz. 

Based on (\ref{eq:W-approximate}), the cavity field $h^{a}=\sum_{j}J_{j}^{a}s_{j}$ can be decomposed into  the sum of the ``signal'' part $h_{\Psi_d}$ and the ``noise'' part $h_{\triangle}^{a}$ as
\begin{align}
h^{a} & =h_{\Psi_d}+h_{\triangle}^{a},\\
h_{\Psi_d} & \doteq\sum_{j\in\Psi_d}\text{\ensuremath{\bar{J}_{j}}}s_{j},\\
h_{\triangle}^{a} & \doteq\frac{1}{\sqrt{N}}\sum_{j}\Delta_{j}^{a}s_{j}\approx\frac{1}{\sqrt{N}}\sum_{j\in\bar{\Psi}_d}\Delta_{j}^{a}s_{j},\label{eq:h-delta}
\end{align}
where the approximation in (\ref{eq:h-delta}) is due to the assumption that there are only finite $\mathcal{O}\left(1\right)$ terms  in $\Psi_d$, which are negligible in
the large system limit, as discussed in Section \ref{subsec:Nearest-Neighours-Effect}. An important consequence of this decomposition is that the signal and noise parts are asymptotically independent as $N$ grows against fixed $d$. This is because as $N$ grows the majority of spins in the noise part become farer and farer from the spins in the NN set $\Psi_d$, and resultantly the dependence vanishes in the limit $N\to\infty$. This asymptotic independence makes the computation feasible in the next subsection.

Moreover, the $\ell_{2}$ norm square of $\boldsymbol{J}^{a}=\left(J^a_j\right)_j$
can be computed as
\begin{align}
\left\Vert \boldsymbol{J}^{a}\right\Vert ^{2} & =\sum_{j\in\Psi_d}\left(\text{\ensuremath{\bar{J}_{j}^{2}}}+2\text{\ensuremath{\bar{J}_{j}}}\frac{\Delta_{j}^{a}}{\sqrt{N}}+\frac{(\Delta_{j}^{a})^{2}}{N}\right)+\frac{1}{N}\sum_{j\in\bar{\Psi}_d}\left(\Delta_{j}^{a}\right)^{2}\nonumber \\
 & 
 \approx 
 \sum_{j\in\Psi_d}\bar{J}_{j}^{2}
 +\frac{1}{N}\left\Vert \boldsymbol{\Delta}^{a}\right\Vert ^{2}_2,\label{eq:wa-l2norm-approx}
\end{align}
where the summation of the noise terms over $\Psi_d$ is ignored in the large system limit, which is again due to the assumption of $\mathcal{O}\left(1\right)$ terms in $\Psi_d$. Hence the signal and noise parts are decoupled again in the regularization term. 

The RSS in (\ref{eq:RSS-def-1}) also takes a simple form: 
\begin{equation}
\mathcal{E} \approx\sum_{j\in\Omega_{1}}\left|J_{j}^{*}-\bar{J}_{j}\right|^{2}+\sum_{j\in\Psi_d\setminus\Omega_{1}}\bar{J}_{j}^{2}+R,\label{eq:RSS-def-1-2}
\end{equation}
where $\Psi_d\setminus\Omega_{1}$ denotes the NN set $\Psi_d$ excluding $\Omega_{1}$, and a macroscopic parameter $R$ is introduced as
\begin{equation}
R=\frac{1}{N}\sum_{j\in\bar{\Psi}_d}\triangle_{j}^{2}, \label{eq:R-sparse-def}
\end{equation}
which indicates the sum of square errors in the set $\bar{\Psi}_d$ and is computed below. 

%%%%%%%%%%%%%%%%%%%%%%%%%%%%%%%%%%%%%%%%%%%%%%%%%%%%%
%%%%%%%%%%%%%%%%%%%%%%%%%%%%%%%%%%%%%%%%%%%%%%%%%%%%%
\subsection{Free Energy Density}
Following the ansatz in Section \ref{subsec:Sparse-Anatz}, we can rewrite the replicated partition function $\left[Z^{n}\right]_{\mathcal{D}^{M}}$ of (\ref{eq:Zn-replicated partition}) as
written as
\begin{align}
&\left[Z^{n}\right]_{\mathcal{D}^{M}} =  \int\prod_{a=1}^{n}d\boldsymbol{J}^{a}e^{-\lambda\beta N \sum_{a=1}^{n}\left\Vert \boldsymbol{J}^{a}\right\Vert ^{2}}\left\{ \sum_{\boldsymbol{s}}P_{\textrm{Ising}}\left(\boldsymbol{s}|\boldsymbol{J}^{*}\right)\exp\left[-\beta\sum_{a=1}^{n}\Phi\left(s_{0}h^{a}\right)\right]\right\} ^{\alpha N}\nonumber \\
 \approx  & \int\prod_{a=1}^{n}d\boldsymbol{\Delta}^{a}e^{-\lambda\beta\left(Nn\sum_{j\in\Psi_d}\bar{J}_{j}^{2}+\sum_{a=1}^{n}\left\Vert \boldsymbol{\Delta}^{a}\right\Vert ^{2}\right)}\times\nonumber \\
 & \left\{ \sum_{\boldsymbol{s}}P_{\textrm{Ising}}\left(\boldsymbol{s}|\boldsymbol{J}^{*}\right)\prod_{a}\int dh_{\triangle}^{a}\delta\left(h_{\triangle}^{a}-\frac{1}{\sqrt{N}}\sum_{j\in\bar{\Psi}_d}\Delta_{j}^{a}s{j}\right)e^{-\beta\sum_{a=1}^{n}\Phi\left(s_{0}\left(\sum_{j\in\Psi_d}\text{\ensuremath{\bar{J}_{j}}}s_{j}+h_{\triangle}^{a}\right)\right)}\right\} ^{\alpha N}\nonumber \\
 =  & \int\prod_{a=1}^{n}d\boldsymbol{\Delta}^{a}e^{-\lambda\beta\left(Nn\sum_{j\in\Psi_d}\bar{J}_{j}^{2}+\sum_{a=1}^{n}\left\Vert \boldsymbol{\Delta}^{a}\right\Vert ^{2}\right)}\times\nonumber \\
 & \left\{ \sum_{s_{0},\boldsymbol{s}_{\Psi_d}}P\left(s_{0},\boldsymbol{s}_{\Psi_d},\left\{ h_{\triangle}^{a}\right\} _{a}|J^{*},\left\{ \boldsymbol{\Delta}^{a}\right\} _{a}\right)e^{-\beta\sum_{a=1}^{n}\Phi\left(s_{0}\left(\sum_{j\in\Psi_d}\text{\ensuremath{\bar{J}_{j}}}s_{j}+h_{\triangle}^{a}\right)\right)}\right\} ^{\alpha N}\nonumber \\
 \approx & \int\prod_{a=1}^{n}d\boldsymbol{\Delta}^{a}e^{-\lambda\beta\left(Nn\sum_{j\in\Psi_d}\bar{J}_{j}^{2}+\sum_{a=1}^{n}\left\Vert \boldsymbol{\Delta}^{a}\right\Vert ^{2}\right)}\times\nonumber \\
 & \left\{ \sum_{s_{0},\boldsymbol{s}_{\Psi_d}}P\left(s_{0},\boldsymbol{s}_{\Psi_d}|J^{*}\right)
 \int
 \prod_{a=1}^{n}dh_{\triangle}^{a}P_{\textrm{cav}}\left(\left\{ h_{\triangle}^{a}\right\} _{a}|\left\{ \boldsymbol{\Delta}^{a}\right\} _{a}\right)e^{-\beta\sum_{a=1}^{n}\Phi\left(s_{0}\left(\sum_{j\in\Psi_d}\text{\ensuremath{\bar{J}_{j}}}s_{j}+h_{\triangle}^{a}\right)\right)}\right\} ^{\alpha N},\label{eq:Zn-sparse-def}
\end{align}
where $\boldsymbol{s}_{\Psi_d}$ is the vector of spins in the NN set $\Psi_d$. In the second line of (\ref{eq:Zn-sparse-def}), $\sum_{j\in\Psi_d}\left(\Delta_{j}^{a}\right)^{2}$
is ignored as in (\ref{eq:wa-l2norm-approx}), and in the last line, the asymptotic independence between $h_{\triangle}^{a}$ and $ \{s_0,\bm{s}_{\Psi_d}\} $ are used. The marginal distribution
$P\left(s_{0},\boldsymbol{s}_{\Psi_d}|J^{*}\right)$ is computed by marginalizing the whole distribution $\sum_{\boldsymbol{s}}P_{\textrm{Ising}}\left(\boldsymbol{s}|\boldsymbol{J}^{*}\right)$
with respect to $\boldsymbol{s}_{\bar{\Psi}_d}$, which can be obtained as
\begin{align}
P\left(s_{0},\boldsymbol{s}_{\Psi_d}|J^{*}\right) & =\sum_{\boldsymbol{s}_{\bar{\Psi}_d}}P\left(\boldsymbol{s}|J^{*}\right). \label{eq:marginal-distribution}
\end{align}
Then, according to the central limit theorem, the noise part $\left\{ h_{\triangle}^{a}\right\} _{a=1}^{n}$
can be regarded as Gaussian variables so that the cavity distribution
$P_{\textrm{cav}}\left(\left\{ h_{\triangle}^{a}\right\} _{a}|\left\{ \boldsymbol{\Delta}^{a}\right\} _{a}\right)$
can be assumed as a multivariate Gaussian distribution. Here we assume the replica symmetry (RS), and hence the following two order parameters are sufficient to characterize the multivariate Gaussian distribution:
\begin{align}
Q\doteq & \frac{1}{N}\sum_{i,j\in\bar{\Psi}_d}\Delta_{i}^{a}C_{ij}^{\backslash0}\Delta_{j}^{a},\label{eq:Q-def}\\
q\doteq & \frac{1}{N}\sum_{i,j\in\bar{\Psi}_d}\Delta_{i}^{a}C_{ij}^{\backslash0}\Delta_{j}^{b},\; \left(a \neq b \right), \label{eq:q-def}
\end{align}
where $\boldsymbol{C}^{\backslash0}=\left(C_{ij}^{\backslash0}\right)_{ij} $
is the correlation matrix of the reduced spin system without $s_{0}$.
% The diagonal elements $\left(C_{ii}^{\backslash0}\right)$ equal to 1 and typically
% $C_{ij}^{\backslash0}=\mathcal{O}\left(1/\sqrt{N}\right),i\neq j$.
As suggested in \cite{bachschmid2017statistical,Abbara2019c},
the non-diagonal elements of $\boldsymbol{C}^{\backslash0}$ will have a nontrivial
contribution and will hence be retained. To write the integration in
terms of the order parameters $Q,q$, we introduce the following
trivial identities:
\begin{align}
1 & =  N\int dQ~\delta\left(\sum_{i,j\neq0}\Delta_{i}^{a}C_{ij}^{\backslash0}\Delta_{j}^{a}-NQ\right),a=1,...,n, \\
1 & = N\int dq~\delta\left(\sum_{i,j\neq0}\Delta_{i}^{a}C_{ij}^{\backslash0}\Delta_{j}^{b}-Nq\right),a<b,a,b\neq*.  \label{eq:trivial-delta-func}
\end{align}
Therefore, $\left[Z^{n}\right]_{\mathcal{D}^{M}}$ can be rewritten as
\begin{align}
& \left[Z^{n}\right]_{\mathcal{D}^{M}}  =e^{-\lambda\beta Nn\sum_{j\in\Psi_d}\bar{J}_{j}^{2}} \int dQdq\int\prod_{a=1}^{n}d\boldsymbol{\Delta}^{a}e^{-\lambda\beta\sum_{a=1}^{n}\left\Vert \boldsymbol{\Delta}^{a}\right\Vert ^{2}}\prod_{a=1}^{n}\delta\left(\sum_{i,j}\Delta_{i}^{a}C_{ij}^{\backslash0}\Delta_{j}^{a}-NQ\right) \times\nonumber \\ & \prod_{a<b}\delta\left(\sum_{i,j}\Delta_{i}^{a}C_{ij}^{\backslash0}\Delta_{j}^{b}-Nq\right)\times\nonumber \\
& \left\{ \sum_{s_{0},\boldsymbol{s}_{\Psi_d}}P\left(s_{0},\boldsymbol{s}_{\Psi_d}|J^{*}\right)
\int
\prod_{a=1}^{n}dh_{\triangle}^{a}P_{\textrm{cav}}\left(\left\{ h_{\triangle}^{a}\right\} _{a}|\left\{ \boldsymbol{\Delta}^{a}\right\} _{a}\right)e^{-\beta\sum_{a=1}^{n}\Phi\left(s_{0}\left(\sum_{j\in\Psi_d}\text{\ensuremath{\bar{J}_{j}}}s_{j}+h_{\triangle}^{a}\right)\right)}\right\} ^{\alpha N} \nonumber \\
& = \int dQdq\exp\left[NS+\alpha N \log L\right],
\end{align}
where
\begin{align}
e^{NS} \doteq & e^{-\lambda\beta Nn\sum_{j\in\Psi_d}\bar{J}_{j}^{2}}\int\prod_{a=1}^{n}d\boldsymbol{\Delta}^{a}e^{-\lambda\beta\sum_{a=1}^{n}\left\Vert \boldsymbol{\Delta}^{a}\right\Vert ^{2}}\prod_{a=1}^{n}\delta\left(\sum_{i,j}\Delta_{i}^{a}C_{ij}^{\backslash0}\Delta_{j}^{a}-NQ\right) \nonumber  \\ 
& \times \prod_{a<b}\delta\left(\sum_{i,j}\Delta_{i}^{a}C_{ij}^{\backslash0}\Delta_{j}^{b}-Nq\right),\label{eq:NS-part-sparse}\\
L  \doteq & \sum_{s_{0},\boldsymbol{s}_{\Psi_d}}P\left(s_{0},\boldsymbol{s}_{\Psi_d}|J^{*}\right)\int\prod_{a=1}^{n}dh_{\triangle}^{a}P_{\textrm{cav}}\left(\left\{ h_{\triangle}^{a}\right\} _{a}|\left\{ \boldsymbol{\Delta}^{a}\right\} _{a}\right)e^{-\beta\sum_{a=1}^{n}\Phi\left(s_{0}\left(\sum_{j\in\Psi_d}\text{\ensuremath{\bar{J}_{j}}}s_{j}+h_{\triangle}^{a}\right)\right)}.\label{eq:L-part-df-1}
\end{align}
After performing some algebraic operations presented in \ref{subsec:Appendix-S} and \ref{subsec:appendix-L}, we obtain the results in the limit
$n\rightarrow0$: 
\begin{align}
\lim_{n\rightarrow0}\frac{S}{n} & =-\lambda\beta\sum_{j\in\Psi_d}\bar{J}_{j}^{2}+\frac{Q\beta}{2}G_{1}^{-1}\left(\beta\left(Q-q\right)\right)-\frac{1}{2}G_{2}\left(G_{1}^{-1}\left(\beta\left(Q-q\right)\right)\right)\nonumber \\
 & +\frac{1}{2}\log\frac{2\pi}{\beta}-\frac{1}{2N}\textrm{Tr}\log\left(\boldsymbol{C}^{\backslash0}\right)^{-1}, \\
\underset{n\rightarrow0}{\lim}\frac{1}{n}\log L & =\sum_{s_{0},\boldsymbol{s}_{\Psi_d}}P\left(s_{0},\boldsymbol{s}_{\Psi_d}|J^{*}\right)\int\mathcal{D}z\log\int\mathcal{D}ve^{-\beta\sum_{a=1}^{n}\Phi\left(s_{0}\left(\sum_{j\in\Psi_d}\text{\ensuremath{\bar{J}_{j}}}s_{j}+\sqrt{Q-q}v+\sqrt{q}z\right)\right)}, 
\end{align}
where $\mathcal{D}z=\frac{dz}{\sqrt{2\pi}}e^{-\frac{z^{2}}{2}}$,
and 
\begin{align}
G_{1}\left(x\right) & =\int d\eta \frac{\rho\left(\eta\right)}{x+2\lambda\eta},\label{eq:G1-x-def-1}\\
G_{2}\left(x\right) & =\int d\eta\rho\left(\eta\right)\log\left(x+2\lambda\eta\right), 
\end{align}
where $\rho\left(\eta\right)$ is the eigenvalue distribution (EVD)
of the inverse correlation matrix, i.e., $\left(\boldsymbol{C}^{\backslash0}\right)^{-1}$,
as shown in \ref{subsec:Eigenvalue-distribution}. Note that
$G_{1}^{-1}$ denotes the inverse function of $G_{1}$, i.e., $x=G_{1}^{-1}\left(y\right)$
implies that $y=G_{1}\left(x\right)$. In the special case of $\lambda=0$,
$G_{1}\left(x\right)=1/x$ and $G_{1}^{-1}\left(y\right)=1/y$. 

Further, we take the limit $\beta\rightarrow\infty$, which requires
the following relation \cite{bachschmid2017statistical,Abbara2019c}: 
\begin{equation}
\lim_{\beta\rightarrow\infty}\beta\left(Q-q\right)=\chi=\mathcal{O}\left(1\right). \label{eq:limit-temperature-condition}
\end{equation}
$\chi=\mathcal{O}\left(1\right)$ is a finite number, and 
according to (\ref{eq:G1-x-def-1}), $G_{1}^{-1}\left(\beta\left(Q-q\right)\right)$
should also be a finite number. Then, denoting $G_{1}^{-1}\left(\beta\left(Q-q\right)\right) \doteq \kappa$,
after performing some algebraic operations, we obtain the free energy density (\ref{eq:F-replica-def})
in the limit $\beta\rightarrow\infty$ as
\begin{align}
f\left(\beta\rightarrow\infty\right) & =-\underset{Q, \kappa, \{\bar{J}_j\}_{j\in \Psi_d}}{\textrm{Extr}}
\Biggl\{
-\lambda\sum_{j\in\Psi_d}\bar{J}_{j}^{2}+\frac{Q}{2}\kappa+\nonumber \\
 & \alpha\sum_{s_{0},\boldsymbol{s}_{\Psi_d}}P\left(s_{0},\boldsymbol{s}_{\Psi_d}|J^{*}\right)\int\mathcal{D}z\underset{y}{\max}\left[-\frac{\left(y-s_{0}\left(\sqrt{Q}z+\sum_{j\in\Psi_d}\text{\ensuremath{\bar{J}_{j}}}s_{j}\right)\right)^{2}}{2G_{1}\left(\kappa\right)}-\Phi\left(y\right)\right] 
 \Biggr\}
 ,\label{eq:limit-free-energy-sparse}
\end{align}
where $\underset{x}{\textrm{Extr}}\left\{ \cdot\right\} $ denotes
extremization w.r.t. $x$. 

%%%%%%%%%%%%%%%%%%%%%%%%%%%%%%%%%%%%%%%%%%%%%%%%%%%%%
%%%%%%%%%%%%%%%%%%%%%%%%%%%%%%%%%%%%%%%%%%%%%%%%%%%%%
\subsection{\label{EOS-subsec}Equations of State (EOS)}
From (\ref{eq:limit-free-energy-sparse}), the extremization condition leads to the following equations of state (EOS):
\begin{align}
\kappa-\frac{\alpha}{\sqrt{Q}}\int\mathcal{D}zz\frac{\partial l\left(y\right)}{\partial y}\mid_{y=\hat{y}} = 0, \\
Q+\alpha G_{1}^{'}\left(\kappa\right)\int\mathcal{D}z\left(\frac{\partial l\left(y\right)}{\partial y}\mid_{y=\hat{y}}\right)^{2} = 0,
\label{eq:EOS-sparse}
\end{align}
where
\begin{align}
\hat{y} & =\underset{y}{\textrm{arg}\max}\left(-\frac{\left(y-s_{0}\left(\sqrt{Q}z+\sum_{j\in\Psi_d}\text{\ensuremath{\bar{J}_{j}}}s_{j}\right)\right)^{2}}{2G_{1}\left(\kappa\right)}-\Phi\left(y\right)\right).\label{eq:y-hat-def}
\end{align}
Moreover, the mean estimates $\text{\ensuremath{\left\{  \bar{J}_{j}\right\} } }_{j\in\Psi_d}$
can also be evaluated by the extremization condition, i.e.,
\begin{align}
0 & =2\lambda\bar{J}_{j}+\alpha\sum_{s_{0},\boldsymbol{s}_{\Psi_d}}P\left(s_{0},\boldsymbol{s}_{\Psi_d}|J^{*}\right)\int\mathcal{D}z\frac{\partial\Phi\left(y\right)}{\partial y}\mid_{y=\hat{y}}s_{0}s_{j},\;j\in\Psi_d,\label{eq:J-bar-condition}
\end{align}
which is a set of linear equations in our case of the quadratic cost function. Note that when the coupling strength is uniform, i.e., $\left|J_{j}^{*}\right|=K,\;j\in\Omega_{1}$,
the strength of the mean estimates $\text{\ensuremath{\bar{J}_{j}}}\in\Omega_{1}$
can also be set to a uniform value $\left|\bar{J}_{j}\right|=\bar{K}=\hat{b}K$, where the bias factor is defined as
\begin{equation}
\hat{b}\doteq\frac{\bar{K}}{K}.\label{eq:bias-def}
\end{equation}
Besides, using the auxiliary variable technique similar to \cite{Abbara2019c}, as shown in \ref{Appendix-MacroParameters-auxiliarymethod}, the macroscopic parameter $R$ in (\ref{eq:R-sparse-def})
can be computed as 
\begin{equation}
R=\frac{1}{N}\sum_{j\in\bar{\Psi}_d}\triangle_{j}^{2}=q\frac{G_{3}^{'}\left(\kappa\right)}{G_{1}^{'}\left(\kappa\right)}, \label{eq:R-def-resuls}
\end{equation}
where 
\begin{equation}
G_{3}\left(x\right)=\int d\eta \frac{\rho\left(\eta\right)\eta}{\left(x+2\lambda\eta\right)},\label{eq:G3-x}
\end{equation}
and $G_{1}^{'}\left(x\right)$ and $G_{3}^{'}\left(x\right)$ are the first-order derivatives of $G_{1}\left(x\right)$ and $G_{3}\left(x\right)$, respectively. Then, given $\text{\ensuremath{\left\{  \bar{J}_{j}\right\} } }_{j\in\Psi_d},\;Q,\;\kappa$, the RSS in (\ref{eq:RSS-def-1}) can be computed as 
\begin{align}
\mathcal{E} & \approx\sum_{j\in\Omega_{1}}\left|J_{j}^{*}-\bar{J}_{j}\right|^{2}+\sum_{j\in\Psi_d\setminus\Omega_{1}}\bar{J}_{j}^{2}+Q\frac{G_{3}^{'}\left(\kappa\right)}{G_{1}^{'}\left(\kappa\right)}.\label{eq:RSS-def-2}
\end{align}
In general, no analytical solution exists for the EOS, but it can be
easily solved using numerical methods, as illustrated in \ref{subsec:Numerical-Solutions}.  

%%%%%%%%%%%%%%%%%%%%%%%%%%%%%%%%%%%%%%%%%%%%%%%%%%%%%
%%%%%%%%%%%%%%%%%%%%%%%%%%%%%%%%%%%%%%%%%%%%%%%%%%%%%
\subsection{\label{subsec:Nearest-Neighours-Effect}Nearest-Neighbor Effect}
In this subsection, we study the NN effect by examining the mean estimates $\text{\ensuremath{\left\{  \bar{J}_{j}\right\} } }_{j\in\Psi_d}$, and the application range of ansatz (\ref{eq:W-approximate}) is also discussed. According to the replica analysis presented above, the mean estimates
$\text{\ensuremath{\left\{  \bar{J}_{j}\right\} } }_{j\in\Psi_d}$ can
be calculated by solving the linear equations
\begin{align}
\left(1+2\lambda/\kappa\right)\bar{J}_{j}+\sum_{i\in\Psi_d,i\neq j}\bar{J}_{i}\left\langle s_{i}s_{j}\right\rangle -\left\langle s_{0}s_{j}\right\rangle =0, \; j\in\Psi_d, \label{eq:J-mean-linear-equation-1}
\end{align}
where $\left\langle s_{i}s_{j}\right\rangle $ is the correlation function w.r.t.
the joint distribution $P\left(s_{0},\boldsymbol{s}_{\Psi_d}|J^{*}\right)$.
First, consider the special case without regularization.
The result is given in Theorem 1. 
%%%%%%%%%%%%%%%%%%%%%%%%%%
\begin{theorem}
For a teacher Ising model with a sparse tree-like coupling network in the paramagnet phase, using linear regression without regularization, the mean estimates $\text{\ensuremath{\left\{  \bar{J}_{j}\right\} } }_{j\in\Psi_d}$
in (\ref{eq:J-mean-linear-equation-1}) are 
\begin{equation}
\bar{J}_{j}=\begin{cases}
\frac{1}{\sum_{k\in\Omega_{1}}\frac{1}{1-\tanh^{2}J_{k}^{*}}-c+1}\cdot\frac{\tanh(J_{j}^{*})}{1-\tanh^{2}(J_{j}^{*})} & j\in\Omega_{1}\\
0, & j\notin\Omega_{1}
\end{cases},\label{eq:solution-Jbar-lamda=00003D0-1}
\end{equation}
where c is the number of first-generation nearest neighbors of $s_{0}$,
i.e., $c=\left|\Omega_{1}\right|$. In particular, with uniform coupling,
i.e., $\left|J_{j}^{*}\right|=K,j\in\Omega_{1}$, it is 
\begin{equation}
\bar{J}_{j}=\begin{cases}
\frac{\tanh K}{1+\left(c-1\right)\tanh^{2}K}\textrm{\rm{sign}}(J_{j}^{*}), & j\in\Omega_{1}\\
0, & j\notin\Omega_{1}
\end{cases}.\label{eq:solution-Jbar-lamda=00003D0}
\end{equation}
\end{theorem}
%%%%%%%%%%%%%%%%%%%%%%%%%%
The proof is given in \ref{subsec:Lemma1-proof}.  Theorem 1 shows that, even under model mismatch, naive linear regression without regularization can reconstruct the active set $\Omega_{1}$, i.e., $\bar{J}_{j}=0,j\in\Psi_d\setminus\Omega_{1}$. This result is consistent with the result in \cite{Abbara2019c},
which is obtained by analyzing the zero-gradient condition for the general loss function. Consequently, when $\lambda=0$, we can simply ignore the biases $\bar{J}_{j},j\notin\Omega_{1}$ of the estimator
when computing the RSS in (\ref{eq:RSS-def-2}). 

However, when there is regularization, the result is different, as stated in Theorem 2.
%%%%%%%%%%%%%%%%%%%%%%%%%%
\begin{theorem}
For a teacher Ising model of uniform coupling strength $K$ with a sparse tree-like coupling network in the paramagnet phase, using $\ell_2$-regularized linear regression with regularization coefficient $\lambda>0$, the mean estimates $\text{\ensuremath{\left\{  \bar{J}_{j}\right\} } }_{j\in\Psi_d}$ in (\ref{eq:J-mean-linear-equation-1}) are biased to nonzero values in the inactive set but decay at least exponentially fast w.r.t. the
distance from spin $s_{0}$ with factor $\delta=\theta\frac{\left(D-1\right)}{D-\theta^{2}}$, where $D=1+\frac{2\lambda}{\kappa}$ and $\theta=\tanh{K}$.
\end{theorem}
%%%%%%%%%%%%%%%%%%%%%%%%%%
The proof is given in \ref{subsec:Lemma2-proof}. Theorem 2 shows that the use of $\ell_2$ regularization in ridge regression introduces biases into the coupling estimates for the inactive set; hence, one must be careful about the potential false positives when using $\ell_2$ regularization.
The biases decay at least exponentially fast w.r.t. the distance $d$ between $s_{j}$ and $s_{0}$. Namely, the relation $\left|\bar{J}_{j}/\bar{J_{k}}\right|<\delta^{d-1}$ holds for $\forall j\in\Omega_{d}$ and the $j$'s ascendant $ k \in\Omega_{1}$. Thus, despite the nonzero biases in the inactive set, the ansatz (\ref{eq:W-approximate}) provides an accurate approximation even if we only consider small finite values of $d$. This result holds for any tree graph, and also applies if the graph is asymptotically tree-like in the large system limit. 

This result can be seen from another perspective. Let us consider a RR graph with uniform coupling strength $K$. The above upper bound to the coupling estimates implies that the bias's total contribution of the $d$-th generation $\Omega_{d}$ to the RSS is also upper bounded as 
\begin{align}
\sum_{j\in \Omega_d}\bar{J}_j^2 & < c\left(c-1\right)^{d-1}\left[\tanh^{2}\left(K\right)\right]^{d-1}\bar{K}_{\Omega_{1}}^2,\label{eq:RSS_NN}
\end{align}
where we use an inequality $\delta=\theta (D-1)/(D-\theta)<\theta=\tanh K$ and denote $\bar{K}_{\Omega_{1}}$ as the absolute value of the mean estimates in $\Omega_{1}$. Thus, from (\ref{eq:RSS_NN}), as long as $\left(c-1\right)\tanh^{2}K<1$, this bias contribution converges to zero as $d$ grows and thus can be ignored when $d$ is large enough. Interestingly, the paramagnetic condition $(c-1)\tanh^2 K<1$~\cite{Abbara2019c,mezard2009information,opper2001advanced} corresponds to this converging condition.

%%%%%%%%%%%%%%%%%%%%%%%%%%%%%%%%%%%%%%%%%%%%%%%%%%%%%
%%%%%%%%%%%%%%%%%%%%%%%%%%%%%%%%%%%%%%%%%%%%%%%%%%%%%
%%%%%%%%%%%%%%%%%%%%%%%%%%%%%%%%%%%%%%%%%%%%%%%%%%%%%
\section{\label{sec:Two-Stage-Estimator}Structure Learning and Two-Stage Estimator}
From the analysis presented in Section \ref{sec:Statistical-mechanical-analysis}, for naive linear regression without regularization, the estimates in the inactive set $\hat{J}_{j} \sim \mathcal{O}(1/\sqrt{N}),\; j\notin\Omega_{1} $ are unbiased. Since the variance of estimator scales as $\mathcal{O}(1/N)$, which is later demonstrated when comparing the theoretical result and numerical experiment,  we can obtain the perfect recovery in the limit $N\to\infty$ by pruning the estimates with an appropriate threshold $K_\textrm{th}(>0)$. It is also possible to show that the probability of successfully screening out false positives approaches one by following the same argument as~\cite{Abbara2019c}. Hence, the structure learning is perfectly achievable in the case without regularization. 

Unfortunately, the naive linear regression is only applicable to the $\alpha > 1$ case, and for the underdetermined region $\alpha < 1$ the regularization is needed. However, the use of $\ell_2$ regularization leads to non-zero biases in the inactive set as stated in Theorem 2, which makes the above pruning method difficult to be successful. To overcome this difficulty, based on the other observation in Theorem 2 that the biases decay at least exponentially fast w.r.t. the distance, we propose a two-stage estimator which combines the advantages of both naive linear regression and $\ell_2$-regularized linear regression. 

The specific procedures of the two-stage estimator are as follows. In the first stage, the $\ell_2$-regularized linear regression is applied and the resultant estimate is denoted as $\boldsymbol{\hat{J}}^{\textrm{stg}1}$. To control false positives, a certain constant threshold value $K_{1}(\sim \mathcal{O}(1))$ is introduced, and the elements of $\boldsymbol{\hat{J}}^{\textrm{stg}1}$ whose absolute values are less than $K_{1}$ are considered as negligible and set to zero, i.e.,
\begin{equation}
\hat{J}_{j}^{\textrm{stg}1\textrm{-th}}=\begin{cases}
\hat{J}_{j}^{\textrm{stg}1}, & \textrm{if}\;\left|\hat{J}_{j}^{\rm{stg}1}\right|>K_{1},\\
0, & \rm{otherwise}.
\end{cases}\label{eq:stage1-threshold}
\end{equation}
In contrast to the above pruning method, the threshold $K_{1}$ is not required to eliminate all the false positives, but it should be sufficiently small to avoid false negatives, which is relatively easy to implement. According to Theorem 2, the biases in the inactive set decay exponentially fast and hence there will be only $\mathcal{O}\left(1\right)$ false positives in $\boldsymbol{\hat{J}}^{\textrm{stg}1\textrm{-th}}$. To further eliminate those $\mathcal{O}\left(1\right)$ false positives, in the second stage, the naive linear regression without regularization is applied only to the support of $\boldsymbol{\hat{J}}^{\textrm{stg}1\textrm{-th}}$, which leads to another estimate $\boldsymbol{\hat{J}}^{\textrm{stg}2}$. We again prune this estimate by introducing another threshold $K_{2}(\sim \mathcal{O}(1))$, which corresponds to $K_{\rm th}$ in the single-step pruning method, to judge the estimate component satisfying $|\hat{{J}}^{\textrm{stg}2}_j|<K_2$ as zero. Since there are only $\mathcal{O}\left(1\right)$ false positives after the first stage, the problem in the second stage effectively corresponds to the situation $\alpha\rightarrow\infty$ ($\mathcal{O}\left(1\right)$ unknowns but with $M=\alpha N \rightarrow\infty$ samples) in the large system limit; hence, the perfect recovery is again possible. These procures provide a practical and reasonable way to achieve the perfect recovery for all $\alpha>0$. We could thus derive the positive answer to the structure learning for inverse Ising problems even in the model mismatch setting. 

In the next section, the effectiveness of the above proposed method is demonstrated in numerical experiments, to show a quantitatively satisfactory performance.

%%%%%%%%%%%%%%%%%%%%%%%%%%%%%%%%%%%%%%%%%%%%%%%%%%%%%
%%%%%%%%%%%%%%%%%%%%%%%%%%%%%%%%%%%%%%%%%%%%%%%%%%%%%
%%%%%%%%%%%%%%%%%%%%%%%%%%%%%%%%%%%%%%%%%%%%%%%%%%%%%
\section{{\label{sec:Numerical-experiments}Numerical experiments}}
Here we conduct numerical experiments to examine the theoretical analysis and the performance of the proposed estimators. The experimental setup is as follows. The teacher Ising model is assumed to have a uniform coupling strength $K$, and the coupling network is assumed to be the RR graph with a connectivity parameter $c$ or the ER graph with the connection probability $p$. As in~\cite{Abbara2019c}, to keep the generated graph sufficiently sparse in the ER case, the probability $p$ is assumed to scale as $p=\bar{c}/N$, yielding the mean degree $\bar{c}$. We assume that the active couplings of the teacher model have the same probability of taking both signs. In addition, $K$ is assumed to be sufficiently small to satisfy the paramagnet assumption of the teacher model \cite{Abbara2019c,mezard2009information,opper2001advanced}. The experimental procedures are similar to those in \cite{Abbara2019c}. First, a random graph is generated and the teacher Ising model with coupling strength $K$ is defined on it. From the teacher model, the spin snapshots are obtained using MC sampling, yielding the dataset $\mathcal{D}^{M}$. Then, we randomly choose a center spin $s_{0}$ from all the spins and infer the associated couplings connected to $s_{0}$ by applying our linear estimators to $\mathcal{D}^{M}$. The experimental values of the macroscopic quantities of interest, such as RSS, can be easily obtained. To obtain the error bars of them, we repeat the sequence of operations many times. Note that in the MC sampling, we started from a random initial configuration and updated the state by the standard Metropolis method; one MC step (MCS) is defined by $N$ trial flips of spins, where $N$ is the total number of spins. We discarded the first $10^{5}$ MCSs as burn-in to avoid systematic errors from the initialization. 

\begin{figure}[thb]
\begin{center}
\includegraphics[width=16cm]{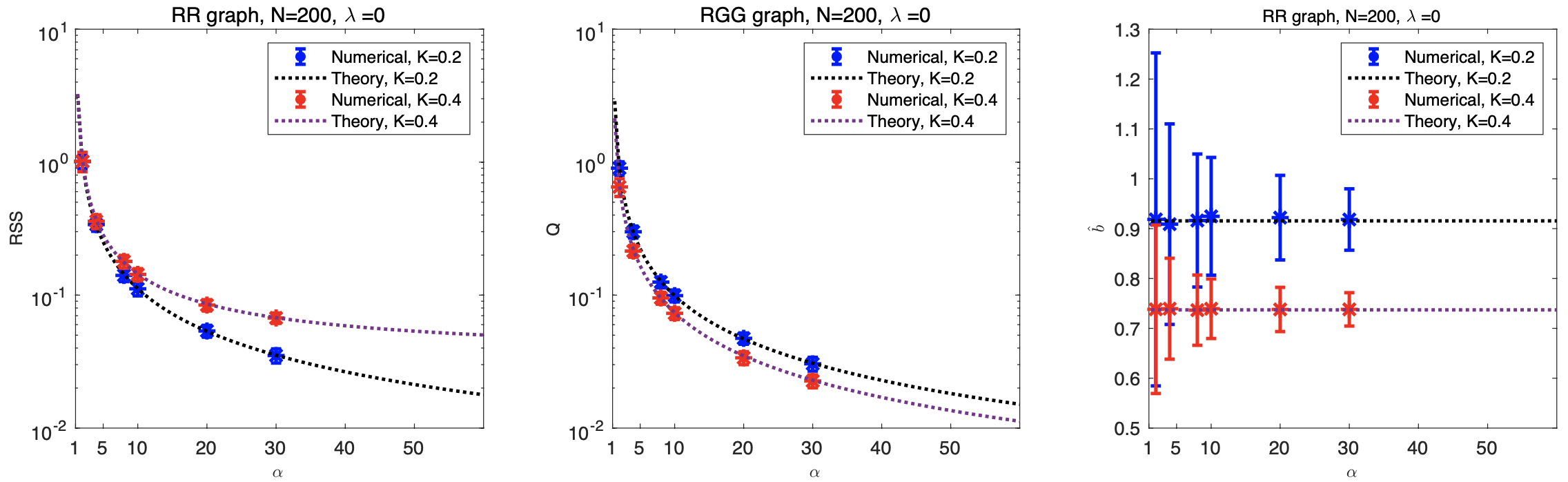}
\caption{Plots of RSS $\mathcal{E}$, $Q$, and bias factor $\hat{b}$ for the RR graph with
$\left(N,c\right)=\left(200,3\right)$ for $K=0.2,0.4$ using linear regression, i.e., $\lambda=0$. The dotted lines and colored markers represent the replica prediction
and numerical values, respectively. The RSS and $Q$ diverge in the limit $\alpha \rightarrow 1$. The error bars are obtained from 100 random runs. \label{fig:Plots-N200-RSSQbias-lambda0}}
\end{center}
\end{figure}

First, let us consider the case without regularization, i.e., $\lambda=0$. Here, the NN set $\Psi_d$ is fixed to be $\Psi_d=\Omega_{1}$ in the theoretical analysis, as the estimates are unbiased in the inactive set as stated in Theorem 1. The theoretical and experimental values of the RSS $\mathcal{E}$, order parameter $Q$, and bias factor $\hat{b}$ for the RR graph are shown in Fig. \ref{fig:Plots-N200-RSSQbias-lambda0}, where the error bars are obtained from 100 random runs. 
%%%%%%%%%%%%%%%%%%%%%
%%%%%%%%%%%%%%%%%%%%%
As can be seen from Fig. \ref{fig:Plots-N200-RSSQbias-lambda0}, the experimental and theoretical results are in fairly good agreement, which supports the validity of the theoretical analysis. The divergence of the RSS and $Q$ at $\alpha \to 1$ corresponds to the phase transition when approaching to the underdetermined region $\alpha< 1$, signaling the limit of using the naive linear regression without regularization.

\begin{figure}[thb]
\begin{center}
\includegraphics[width=15cm]{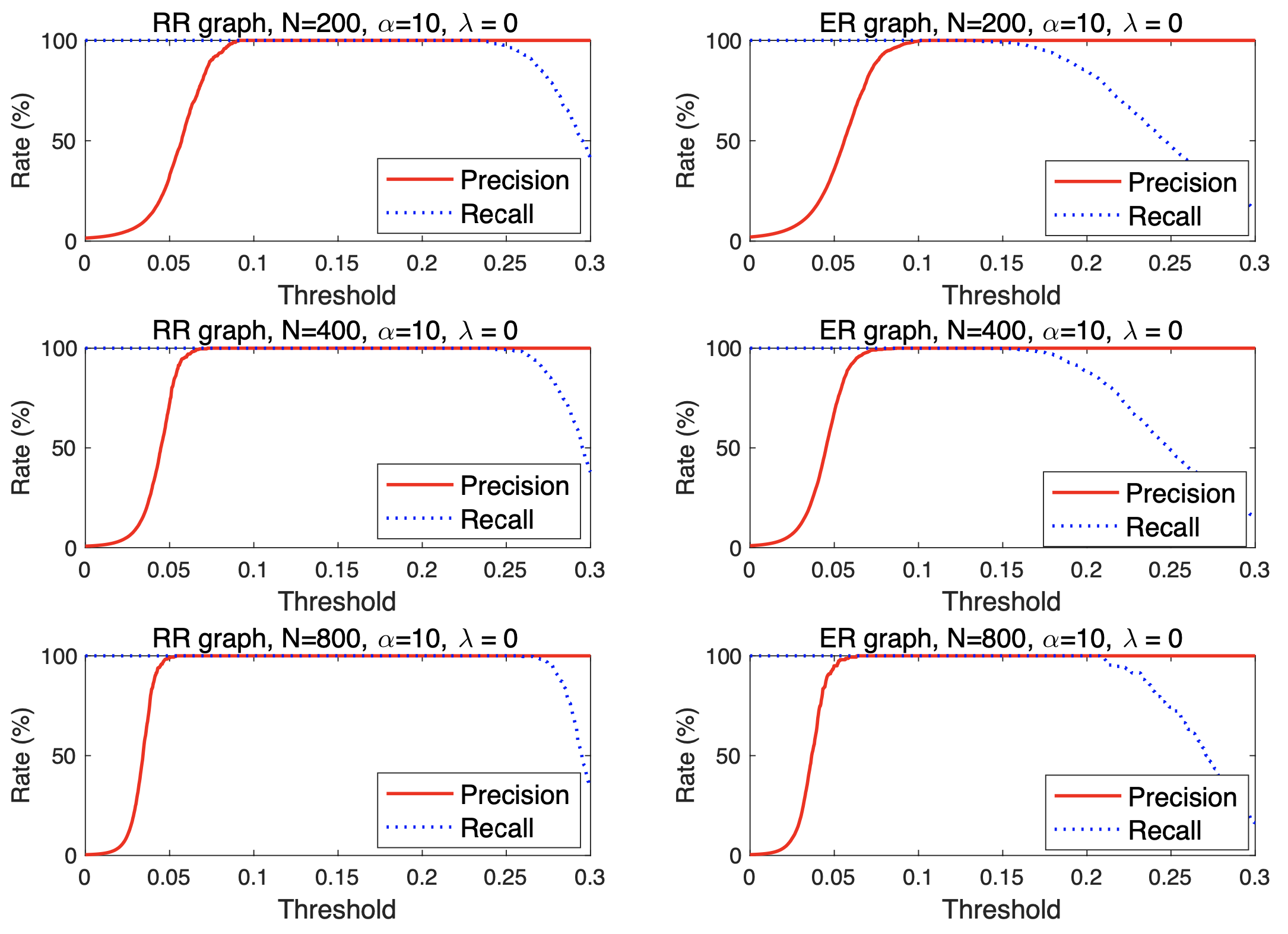}
\caption{Precision and Recall plotted against the threshold $K_{\rm th}$ for RR and ER graphs using naive linear regression without regularization for different $N$ in the case of $K=0.4$ and $\alpha=10$. For the RR graph, $c=3$; for the ER graph, $\bar{c}=4$. The dotted and solid lines represent Recall and Precision,
respectively. Ten different ER graphs are generated, each with two independent MC samplings, and learning is then conducted for all $i=0, ..., N-1$. 
\label{fig:RRvsER_noL2}}
\end{center}
\end{figure}

To see the structure learning performance, Fig. \ref{fig:RRvsER_noL2} shows the empirical values of Recall and Precision defined in \Req{Precision-Recall} for both the RR and ER graphs when $\alpha=10$ (results with other values of $\alpha > 1$ are similar). 
%%%%%%%%%%%%%%%%%%%%%
%%%%%%%%%%%%%%%%%%%%%
Perfect recovery is achieved when both Recall and Precision are equal to 1, and we can see there exists a threshold interval actually realizing this. As $N$ increases, this threshold interval becomes larger, and, as our theoretical analysis indicates, it should be $\left(0,\bar{K}_{\Omega_1} \right)$ in the large system limit, where $\bar{K}_{\Omega_1} = \underset{j\in\Omega_{1}}{\min}\left|\bar{J}_{j}\right|$ is the minimum mean estimate in the active set $\Omega_{1}$. This sufficiently wide interval makes the use of naive linear regression practical. 

\begin{figure}[thb]
\begin{center}
\includegraphics[width=16cm]{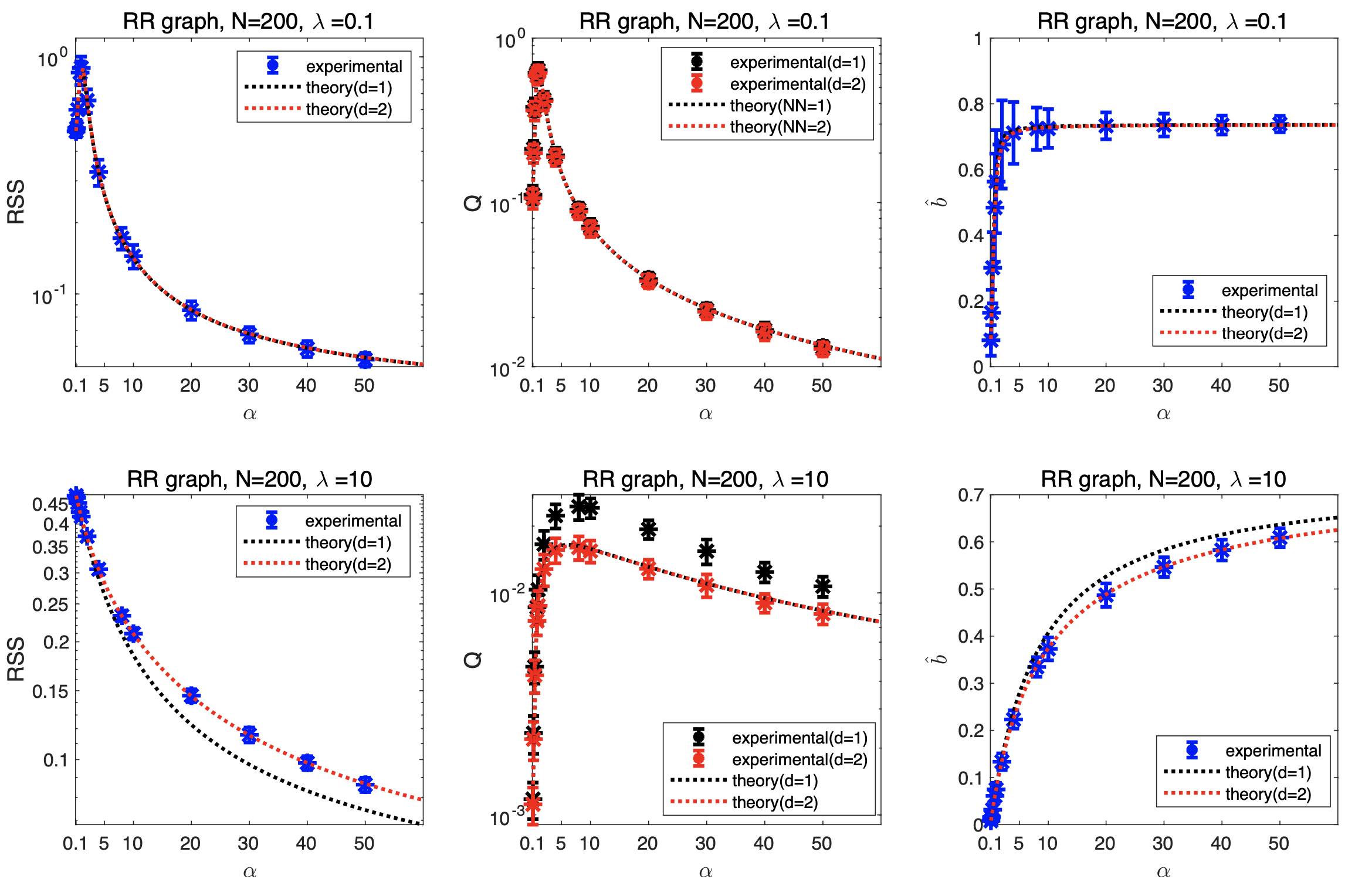}
\caption{Plots of RSS $\mathcal{E}$, $Q$ , and bias factor $\hat{b}$ for the RR graph with
$\left(N,c\right)=\left(200,3\right)$ with $K=0.4$ using ridge regression with $\lambda = 0.1,10$. The dotted lines and colored markers represent the replica prediction and numerical values, respectively. Note that the experimental values of Q are different for $d=1$ and $d=2$ since $Q$ is related to the definition of $\Psi_d$, as shown in (\ref{eq:Q-def}). The error bars are obtained from 100 random runs.  
\label{fig:Plots-N200-RSSQbias}}
\end{center}
\end{figure}

\begin{figure}[thb]
\begin{center}
\includegraphics[width=16cm]{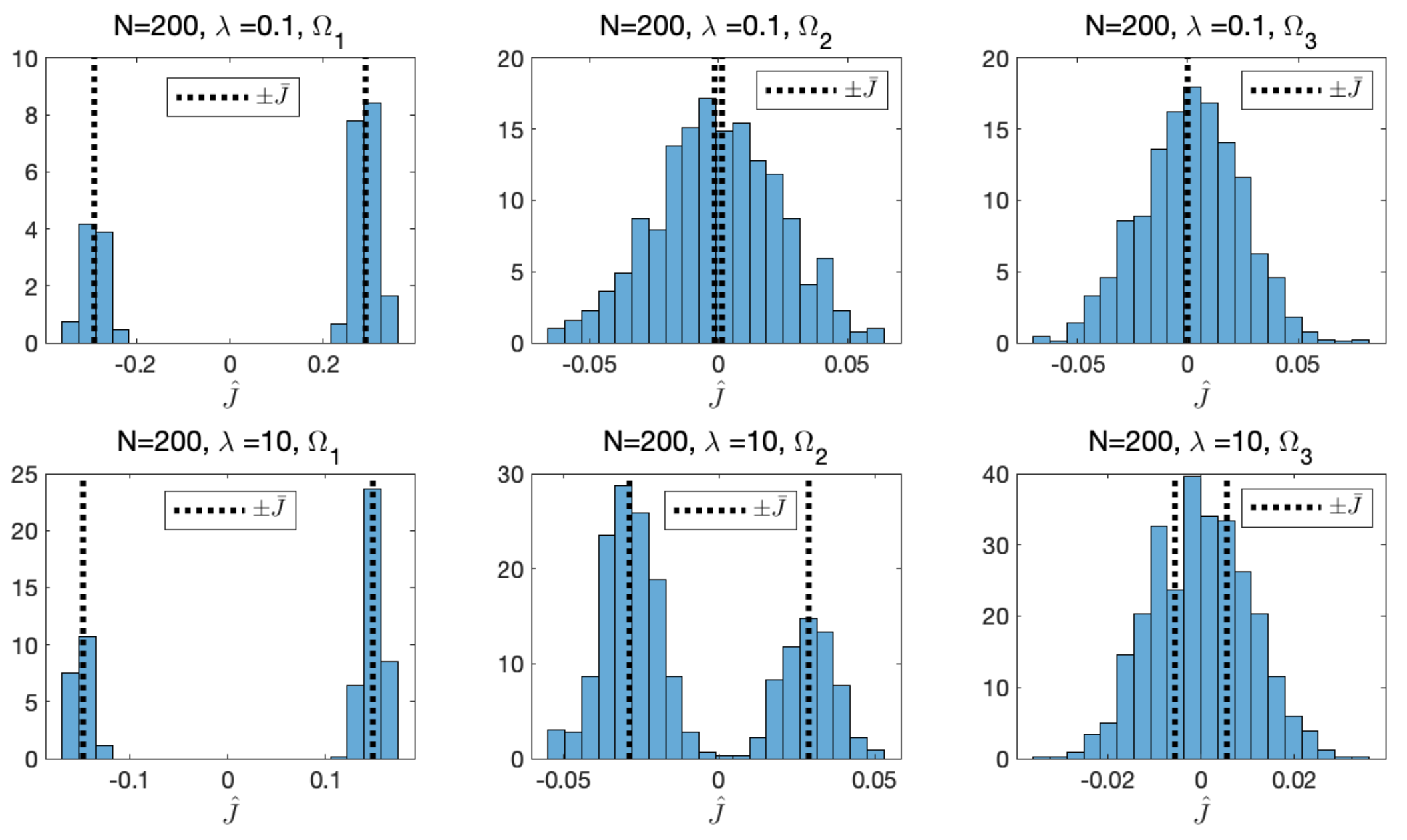}
\caption{Histograms of the estimations $\boldsymbol{\hat{J}}$ in the first three generations $\Omega_d,\; d=1,2,3$ from $s_{0}$. The
system parameters are $\left(N,K,c,\alpha,\right)=\left(200,400,3,10\right)$ with $\lambda = 0.1, 10$. The histograms are generated from 100 random runs. In contrast to linear regression, there are nonzero biases in the NN inactive spins $\Omega_{d}, d\geq2$, which cannot always be ignored, especially when $\lambda$ and/or $K$ is large, e.g., the histograms of the inactive couplings in $\Omega_2$ are far from the zero-mean Gaussian when $\lambda=10$, as shown in the lower part. 
\label{fig:Distr-N200-K400-lamda10}}
\end{center}
\end{figure}

Next, we turn to the finite regularization or the ridge regression case $\lambda>0$. The theoretical and experimental values of the RSS $\mathcal{E}$, order parameter $Q$, and bias factor $\hat{b}$ for the RR graph are shown in Fig. \ref{fig:Plots-N200-RSSQbias}. 
%%%%%%%%%%%%%%%%%%%%%
%%%%%%%%%%%%%%%%%%%%%
Compared to Fig. \ref{fig:Plots-N200-RSSQbias-lambda0}, there are three main differences. First, the use of $\ell_2$ regularization successfully eliminates the divergence in the limit $\alpha \rightarrow 1$, making it applicable in the underdetermined region $\alpha < 1$. Second, the biases of the neighboring spins cannot always be ignored, especially when $\lambda$ and/or $K$ is large, as indicated by the lower part of Fig.  \ref{fig:Plots-N200-RSSQbias}, which shows the apparent discrepancy between the experimental results and the theoretical prediction when ignoring all the biases $\text{\ensuremath{\left\{  \bar{J}_{j}\right\} } }_{j\in\Omega_d,d \geq 2}$. This implies that one must be careful about the potential false positives caused by the nonzero biases on the $d\geq 2$ generations when using $\ell_2$-regularized linear regression. This is consistent with the result in Theorem 2:  when $\lambda$ and/or $K$ is large, the decay factor $\delta=\theta\frac{\left(D-1\right)}{D-\theta^{2}}$ is high; hence, the biases in the inactive set decay slowly. Yet, owing to the exponential decay, it is considered to be possible to make a good approximation just by choosing some small  value of $d$. Actually as shown in Fig. \ref{fig:Plots-N200-RSSQbias}, putting $d=2$ leads to fairly good agreement between the theoretical and experimental results. This can also be verified by empirically evaluating the distribution of estimates $\text{\ensuremath{\left\{  \hat{J}_{j}\right\} } }$ in different $\Omega_d$, which is shown in Fig. \ref{fig:Distr-N200-K400-lamda10} for the first three generations. 
%%%%%%%%%%%%%%%%%%%%%
%%%%%%%%%%%%%%%%%%%%%
In the lower part of Fig. \ref{fig:Distr-N200-K400-lamda10}, when the regularization coefficient is $\lambda=10$, the histograms of the inactive couplings in $\Omega_{d}$ with $d=2$ are far from the zero-mean Gaussian. Therefore, in this case, apart from the true active set $\Omega_{1}$, the NN spins in $\Omega_{2}$ should also be considered as indicated in Fig. \ref{fig:Plots-N200-RSSQbias}. The dashed straight line represents the mean estimates $\text{\ensuremath{\left\{  \bar{J}_{j}\right\} } }$ with different distances $d$ from $s_{0}$ computed from (\ref{eq:J-bar-condition}). When the regularization coefficient is small, e.g., $\lambda=0.1$, the histograms of inactive couplings in $\Omega_{d},\;d\geq2$ are similar to the zero-mean Gaussian; see the upper part of Fig. \ref{fig:Distr-N200-K400-lamda10}. In this case, ignoring the spins with $d\geq2$ in the theoretical analysis still leads to good agreement with the experimental result as shown in Fig. \ref{fig:Plots-N200-RSSQbias}. The last difference between $\ell_2$-regularized linear regression and naive linear regression is that the bias factor $\hat{b}$ is not a constant; it increases as $\alpha$ increases and $\lambda$ decreases.

\begin{figure}[thb]
\begin{center}
\includegraphics[width=15cm]{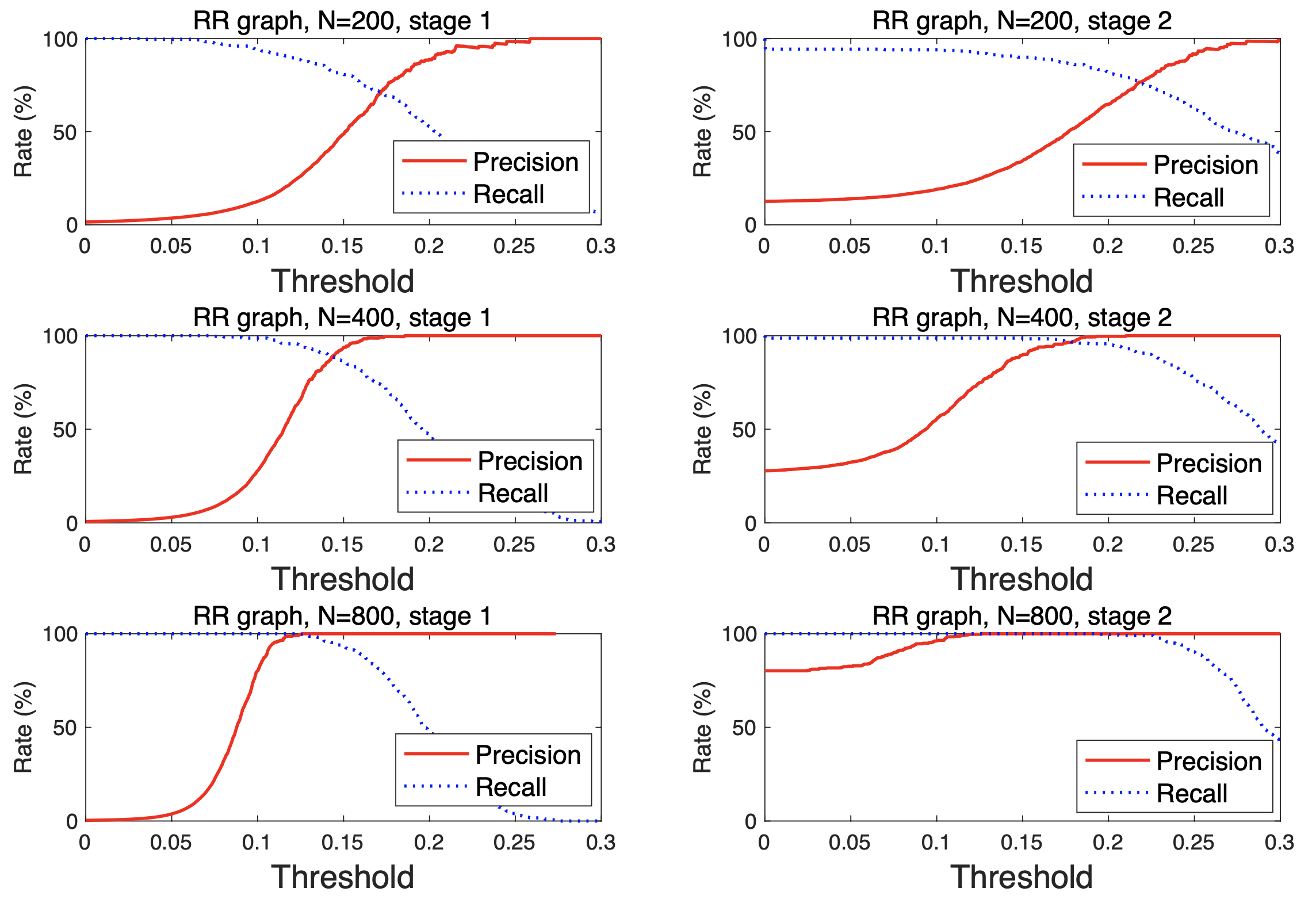}
\caption{Precision and Recall for the RR graph using the two-stage estimator
for different $N$ in the case of $K=0.4,\lambda=0.1$, and $\alpha=0.8$.
The dotted and solid lines represent Recall and Precision, respectively. 
The left panels are for the first stage and are plotted against $K_{1}$, while the right ones are for the second stage and are plotted against $K_2$ after the first stage pruning with $K_{1}=0.1$. The results are obtained from 100 random runs. 
\label{fig:two-stage-recall-precision}}
\end{center}
\end{figure}

\begin{figure}[htb]
\begin{center}
\includegraphics[width=16cm]{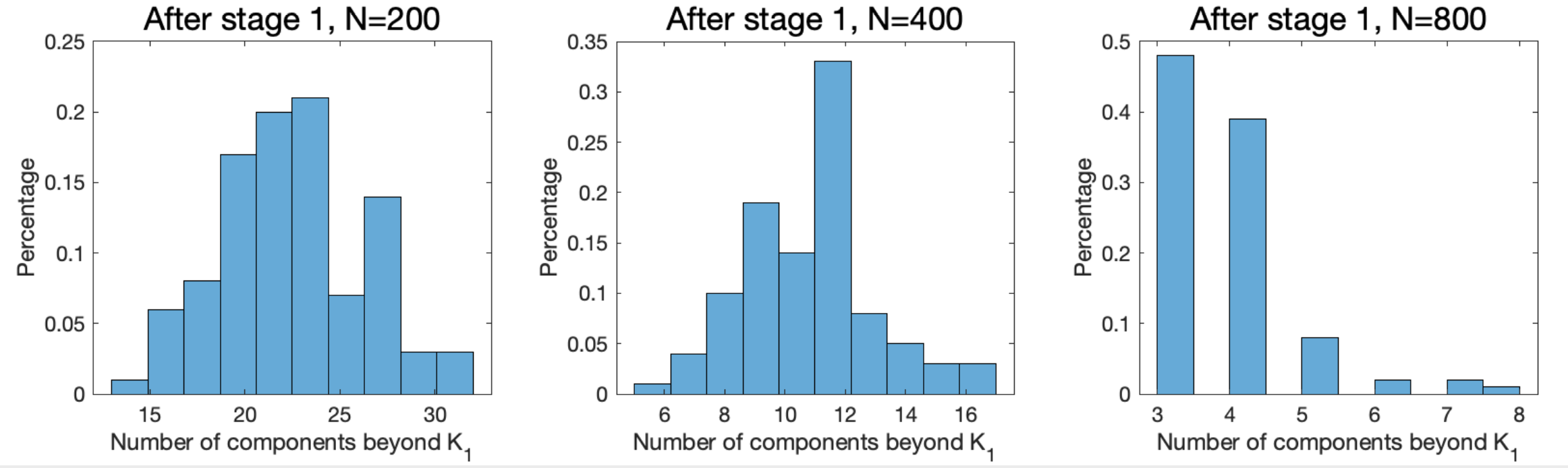}
\caption{Histogram of the number of components selected after the first stage in the
two-stage linear estimator for RR graph of different $N$ in the case of $K=0.4,\lambda=0.1$,
and $\alpha=0.8$. For a fixed threshold $K_{1}=0.1$, as $N$ increases, the number of components beyond the threshold $K_{1}$ decreases as the noise variance scales as $\mathcal{O}\left(1/N\right)$. The results are obtained from 100 random runs. 
\label{fig:two-stage-beyond-threshold}}
\end{center}
\end{figure}

\begin{figure}[htb]
\begin{center}
\includegraphics[width=15cm]{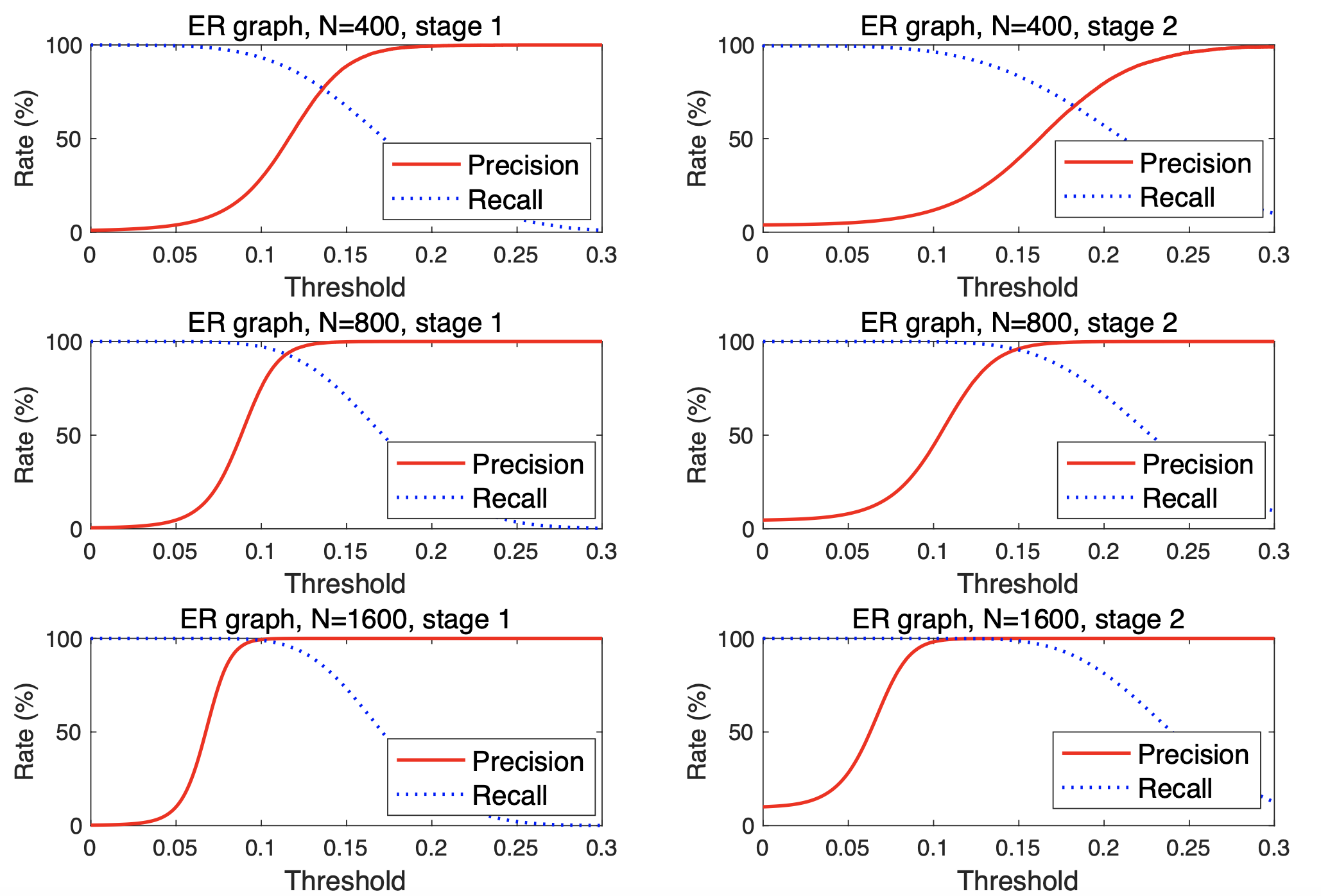}
\caption{Precision and Recall for the ER graph using the two-stage estimator for different $N$ in the case of $K=0.4,\;\lambda=0.1,\; \bar{c}=4$,  and $\alpha=0.9$. The dotted and solid lines represent Recall and Precision, respectively. The left panels are for the first stage and are plotted against $K_{1}$, while the right ones are for the second stage and are plotted against $K_2$ after the first stage pruning with $K_{1}=0.05$. Owing to the finite size effect, exact perfect recovery is not achieved (though when $N=1600$  99.90\% empirical Recall and 99.90\% empirical Precision can be achieved simultaneously), but the improving tendency as $N$ increases is observed, implying perfect recovery in the limit $N \to \infty$. Ten different ER graphs are generated, each with two independent MC samplings, and learning is then conducted for all $i=0, ..., N-1$. 
\label{fig:ERtwo-stage-recall-precision}}
\end{center}
\end{figure}

Finally, the effectiveness of the proposed two-stage linear estimator for structure learning is evaluated in the case of the RR and ER graphs. Fig. \ref{fig:two-stage-recall-precision} shows
a typical result of the empirical Precision and Recall for the RR graph using the two-stage linear estimator with different $N$ when $K=0.4,\;\alpha=0.8,\;\lambda=0.1$.
%%%%%%%%%%%%%%%%%%%%%
%%%%%%%%%%%%%%%%%%%%%
Perfect structure recovery can be achieved with a properly chosen threshold as $N$ increases, e.g., as seen for $N=800$ in Fig. \ref{fig:two-stage-recall-precision}, which verifies the analysis in Section \ref{sec:Two-Stage-Estimator}. It is worth noting that, since the noise variance scales as $\mathcal{O}(1/N)$, the number of components beyond a certain threshold decreases as $N$ increases, as shown in Fig. \ref{fig:two-stage-beyond-threshold}. 
%%%%%%%%%%%%%%%%%%%%%
%%%%%%%%%%%%%%%%%%%%%
This means that as long as the threshold $K_1 \sim \mathcal{O}(1)$ is chosen to be sufficiently small to avoid ignoring true positives, the number of false positives after the first stage can be reduced to a certain $\mathcal{O}(1)$ value as $N \to \infty$. This effectively yields the asymptotic limit of $M \to \infty$ keeping $N$ $\mathcal{O}(1)$; thus, one can easily distinguish true positives from false positives in the second stage. The validity of the two-stage estimator is also evaluated in the case of the ER graph with mean degree $\bar{c}=4$ when $K=0.4,\;\alpha=0.9,\;\lambda=0.1$, as shown in Fig. \ref{fig:ERtwo-stage-recall-precision}. 
%%%%%%%%%%%%%%%%%%%%%
%%%%%%%%%%%%%%%%%%%%%
Although the perfect recovery is not completely achieved (when $N=1600$, there is a threshold interval where empirical Recall and Precision both achieve 99.90\% simultaneously.) owing to the finite size effect, the empirical result in Fig. \ref{fig:ERtwo-stage-recall-precision} indicates the tendency of improvement as $N$ increases, which implies perfect recovery for $N \to \infty$. Note that in the case of the ER graph, we generated 10 different graphs, each with two independent MC samplings, and then conducted learning for all $i=0, ..., N-1$.

%%%%%%%%%%%%%%%%%%%%%%%%%%%%%%%%%%%%%%%%%%%%%%%%%%%%%
%%%%%%%%%%%%%%%%%%%%%%%%%%%%%%%%%%%%%%%%%%%%%%%%%%%%%
%%%%%%%%%%%%%%%%%%%%%%%%%%%%%%%%%%%%%%%%%%%%%%%%%%%%%
\section{\label{sec:Summary-and-discussion}Summary and Discussion}
In this paper, we theoretically investigated the performance of the $\ell_2$- regularized linear estimator applied to the inverse Ising problem in the teacher-student scenario where the couplings of the teacher network are assumed to be sparse and the student has no prior knowledge of its structure and associated parameters, as a representative model mismatch situation. A special focus is on the reconstruction performance of the teacher coupling network. Using the replica and cavity methods of statistical mechanics, we showed that despite the model mismatch, one can perfectly reconstruct the network structure. This is naturally realized owing to the unbiasedness of the linear estimator in the inactive set when the regularization is absent, while it is efficiently achieved using the proposed two-stage estimator when the regularization is present. The proposed two-stage estimator is applicable even when the dataset size is smaller than the number of spins. The results of experiments conducted on locally tree-like graphs \cite{bollobas2001random,dembo2010ising} verified the validity of both the theoretical analysis and the effectiveness of the linear estimator in structure learning in inverse Ising problems. 

The two critical assumptions in this study are the ansatz for handling the cavity field (\ref{eq:W-approximate}) and the paramagnetic assumption for the teacher network. As discussed in Section \ref{subsec:Sparse-Anatz}, the ansatz holds for trees and asymptotic tree-like graphs. The paramagnetic assumption implies that the coupling strength should be sufficiently small. These assumptions restrict the applicability of the presented result, and thus overcoming such limitations will be an important direction for future work. 

Another important direction we think is the use of the $\ell_1$ regularization. This regularization is more popular in sparse estimation and also has been studied in inverse Ising problems~\cite{aurell2012inverse, decelle2014pseudolikelihood, ravikumar2010high,santhanam2012information, bresler2015efficiently}. Unfortunately, this regularization breaks the rotational symmetry of the coupling vector and hence the present analysis cannot be directly applied. It is necessary to invent additional theoretical techniques to overcome this, and such theoretical efforts are currently undergoing.

%%%%%%%%%%%%%%%%%%%%%%%%%%%%%%%%%%%%%%%%%%%%%%%%%%%%%
%%%%%%%%%%%%%%%%%%%%%%%%%%%%%%%%%%%%%%%%%%%%%%%%%%%%%
%%%%%%%%%%%%%%%%%%%%%%%%%%%%%%%%%%%%%%%%%%%%%%%%%%%%%
\section*{Acknowledgement}
This work was supported by JSPS KAKENHI Nos. 17H00764, 18K11463, and 19H01812, and JST CREST Grant Number JPMJCR1912, Japan.

%%%%%%%%%%%%%%%%%%%%%%%%%%%%%%%%%%%%%%%%%%%%%%%%%%%%%
%%%%%%%%%%%%%%%%%%%%%%%%%%%%%%%%%%%%%%%%%%%%%%%%%%%%%
%%%%%%%%%%%%%%%%%%%%%%%%%%%%%%%%%%%%%%%%%%%%%%%%%%%%%

\appendix
%%%%%%%%%%%%%%%%%%%%%%%%%%%%%%%%%%%%%%%%%%%%%%%%%%%%%
%%%%%%%%%%%%%%%%%%%%%%%%%%%%%%%%%%%%%%%%%%%%%%%%%%%%%
%%%%%%%%%%%%%%%%%%%%%%%%%%%%%%%%%%%%%%%%%%%%%%%%%%%%%
\section*{Appendices}
\addcontentsline{toc}{section}{Appendices}
\renewcommand{\thesubsection}{Appendix \Alph{subsection}\noindent}

%%%%%%%%%%%%%%%%%%%%%%%%%%%%%%%%%%%%%%%%%%%%%%%%%%%%%
%%%%%%%%%%%%%%%%%%%%%%%%%%%%%%%%%%%%%%%%%%%%%%%%%%%%%
\subsection{{\label{subsec:Appendix-S}Computation of $e^{NS}$}}
According to the definition in (\ref{eq:NS-part-sparse}), we have 
\begin{align}
e^{NS} = &e^{-\lambda\beta Nn\sum_{j\in\Psi_d}\bar{J}_{j}^{2}}\int\prod_{a=1}^{n}d\boldsymbol{\Delta}^{a}e^{-\lambda\beta\sum_{a=1}^{n}\left\Vert \boldsymbol{\Delta}^{a}\right\Vert ^{2}}\prod_{a=1}^{n}\delta\left(\sum_{i,j}\Delta_{i}^{a}C_{ij}^{\backslash0}\Delta_{j}^{a}-NQ\right) \nonumber  \\
& \times \prod_{a<b}\delta\left(\sum_{i,j}\Delta_{i}^{a}C_{ij}^{\backslash0}\Delta_{j}^{b}-Nq\right). \label{eq:NS-part-def-appendix}
\end{align}
The non-diagonality of $\boldsymbol{C}^{\backslash0}=\left\{ C_{ij}^{\backslash0}\right\} $
will complicate subsequent computations; hence, we first diagonalize
it by introducing an orthogonal matrix $U$ such that $\boldsymbol{C}^{\backslash0}=U^{T}\Lambda U$,
where $\Lambda=\rm{diag}\left[\gamma_{1},\ldots,\gamma_{N-1}\right]$.
Consequently, the term $\sum_{i,j}\Delta_{i}^{a}C_{ij}^{\backslash0}\Delta_{j}^{a}$
becomes
\begin{align}
\sum_{i,j}\Delta_{i}^{a}C_{ij}^{\backslash0}\Delta_{j}^{a} & =\left(\boldsymbol{\triangle}^{a}\right)^{T}\boldsymbol{C}^{\backslash0}\boldsymbol{\triangle}^{a}\nonumber \\
 & =\left(\boldsymbol{\triangle}^{a}\right)^{T}U^{T}\Lambda U\boldsymbol{\triangle}^{a}\nonumber \\
 & =\left(U\boldsymbol{\triangle}^{a}\right)^{T}\Lambda\left(U\boldsymbol{\triangle}^{a}\right)\nonumber \\
 & =\left(\boldsymbol{\tilde{\triangle}}^{a}\right)^{T}\Lambda\boldsymbol{\tilde{\triangle}}^{a},\label{eq:variablechangeU}
\end{align}
where $\boldsymbol{\tilde{\triangle}}^{a}=U\boldsymbol{\triangle}^{a}$.
Similarly, $\sum_{i,j}\Delta_{i}^{a}C_{ij}^{\backslash0}\Delta_{j}^{b}=\left(\boldsymbol{\tilde{\triangle}}^{a}\right)^{T}\Lambda\boldsymbol{\tilde{\triangle}}^{b}$,
and $\left\Vert \boldsymbol{\tilde{\triangle}}^{a}\right\Vert ^{2}=\left\Vert \boldsymbol{\triangle}^{a}\right\Vert ^{2}$.
% Since $U$ is orthogonal matrix, then $d\boldsymbol{\Delta}^{a}=d{\tilde{\triangle}}^{a}$.
By performing the variable transformation in (\ref{eq:NS-part-def-appendix}) and denoting $\boldsymbol{\tilde{\triangle}}$ as $\boldsymbol{\triangle}$,
we obtain
\begin{align}
e^{NS} \doteq & e^{-\lambda\beta Nn\sum_{j\in\Psi_d}\bar{J}_{j}^{2}}\int\prod_{a=1}^{n}d\boldsymbol{\Delta}^{a}e^{-\lambda\beta\sum_{a=1}^{n}\left\Vert \boldsymbol{\Delta}^{a}\right\Vert ^{2}}\nonumber \\
 & \times\prod_{a=1}^{n}\delta\left(\left(\boldsymbol{\Delta}^{a}\right)^{T}\Lambda\boldsymbol{\Delta}^{a}-NQ\right)\prod_{a<b}\delta\left(\left(\boldsymbol{\Delta}^{a}\right)^{T}\Lambda\boldsymbol{\Delta}^{b}-Nq\right).\label{eq:NS-afterOrthogonize}
\end{align}
Then, the delta functions can be expressed as integrals over auxiliary
parameters using the Fourier transform of the delta function, i.e., 
\begin{equation}
\begin{cases}
\delta\left(\left(\boldsymbol{\Delta}^{a}\right)^{T}\Lambda\boldsymbol{\Delta}^{a}-NQ\right)=\int d\hat{Q}e^{\hat{Q}\left(\left(\boldsymbol{\Delta}^{a}\right)^{T}\Lambda\boldsymbol{\Delta}^{a}-NQ\right)},\\
\delta\left(\left(\boldsymbol{\Delta}^{a}\right)^{T}\Lambda\boldsymbol{\Delta}^{b}-Nq\right)=\int d\hat{q}e^{\hat{q}\left(\left(\boldsymbol{\Delta}^{a}\right)^{T}\Lambda\boldsymbol{\Delta}^{b}-Nq\right)},
\end{cases}\label{eq:delta-Foulier}
\end{equation}
where the integration over $\hat{Q},\hat{q}$ is on the imaginary
axis. Hence, (\ref{eq:NS-afterOrthogonize}) can be rewritten
as
\begin{align}
e^{NS} = & \int d\hat{Q}d\hat{q}e^{-N\left(n\hat{Q}Q+\frac{n(n-1)}{2}\hat{q}q\right)}e^{-\lambda\beta Nn\sum_{j\in\Psi_d}\bar{J}_{j}^{2}}\int\prod_{a=1}^{n}d\boldsymbol{\Delta}^{a}e^{-\lambda\beta\sum_{a=1}^{n}\left\Vert \boldsymbol{\Delta}^{a}\right\Vert ^{2}}\nonumber \\
 & \times\exp\left\{ \hat{Q}\sum_{a}\left(\boldsymbol{\Delta}^{a}\right)^{T}\Lambda\boldsymbol{\Delta}^{a}+\hat{q}\sum_{a<b}\left(\boldsymbol{\Delta}^{a}\right)^{T}\Lambda\boldsymbol{\Delta}^{b}\right\} ,\nonumber \\
  = & \int d\hat{Q}d\hat{q}e^{-N\left(n\hat{Q}Q+\frac{n(n-1)}{2}\hat{q}q\right)}e^{-\lambda\beta N\sum_{j\in\Psi_d}\bar{J}_{j}^{2}}\int\prod_{a=1}^{n}d\boldsymbol{\Delta}^{a}e^{-\lambda\beta\sum_{a=1}^{n}\left\Vert \boldsymbol{\Delta}^{a}\right\Vert ^{2}}\nonumber \\
 & \times\exp\left\{ \left(\hat{Q}-\frac{\hat{q}}{2}\right)\sum_{a}\left(\boldsymbol{\Delta}^{a}\right)^{T}\Lambda\boldsymbol{\Delta}^{a}+\frac{\hat{q}}{2}\sum_{a,b}\left(\boldsymbol{\Delta}^{a}\right)^{T}\Lambda\boldsymbol{\Delta}^{b}\right\} ,\nonumber \\
 & =\int d\hat{Q}d\hat{q}e^{S_{X}}\int\prod_{a=1}^{n}d\boldsymbol{\Delta}^{a}e^{U},\label{eq:eSx+eU}
\end{align}
where
\begin{align}
S_{X} & \doteq-N\left(\lambda\beta n\sum_{j\in\Psi_d}\bar{J}_{j}^{2}+n\hat{Q}Q+\frac{n(n-1)}{2}\hat{q}q\right),\label{eq:Sx}\\
U \doteq & -\lambda\beta\sum_{a}\left\Vert \boldsymbol{\Delta}^{a}\right\Vert ^{2}\nonumber \\
 & 
 + \left(\hat{Q}-\frac{\hat{q}}{2}\right)\sum_{a}\left(\boldsymbol{\Delta}^{a}\right)^{T}\Lambda\boldsymbol{\Delta}^{a}+\frac{\hat{q}}{2}\sum_{a,b}\left(\boldsymbol{\Delta}^{a}\right)^{T}\Lambda\boldsymbol{\Delta}^{b}.\label{eq:U-def}
\end{align}
Note that in (\ref{eq:U-def}), different replicas $\boldsymbol{\Delta}^{a}$,
$\boldsymbol{\Delta}^{b}$ are coupled with each other, which makes
it difficult to compute the integration. To overcome this problem,
the Hubbard--Stratonovich transformation is used, i.e.,
\begin{equation}
e^{\frac{cy^{2}}{2}}=\frac{1}{\sqrt{2\pi c}}\int e^{-\frac{x^{2}}{2c}+xy}dx.\label{eq:H-S transform}
\end{equation}
To apply it, we rewrite the term $\sum_{a,b}\left(\boldsymbol{\Delta}^{a}\right)^{T}\Lambda\boldsymbol{\Delta}^{b}$
as
\begin{equation}
\sum_{a,b}\left(\boldsymbol{\Delta}^{a}\right)^{T}\Lambda\boldsymbol{\Delta}^{b}=\sum_{a,b}\sum_{i}\gamma_{i}\triangle_{i}^{a}\triangle_{i}^{b}=\sum_{i}\gamma_{i}\left(\sum_{a}\triangle_{i}^{a}\right)^{2}
\end{equation}
so that
\begin{align}
e^{\frac{\hat{q}}{2}\sum_{a,b}\left(\boldsymbol{\triangle}^{a}\right)^{T}\Lambda\boldsymbol{\triangle}^{b}} & =\prod_{i}e^{\frac{\gamma_{i}\hat{q}\left(\sum_{a}\triangle_{i}^{a}\right)^{2}}{2}}\nonumber \\
 & =\prod_{i}\int\frac{dz_{i}}{\sqrt{2\pi}}e^{-\frac{z_{i}^{2}}{2}+\sqrt{\gamma_{i}\hat{q}}z_{i}\left(\sum_{a}\triangle_{i}^{a}\right)} \nonumber \\
 & =\prod_{i}\int\mathcal{D}z_{i}e^{\sqrt{\gamma_{i}\hat{q}}z_{i}\left(\sum_{a}\triangle_{i}^{a}\right)},\label{eq:H-S-result}
\end{align}
where the change of variable $x_{i}=\sqrt{\gamma_{i}\hat{q}}z_{i}$
is applied and $\mathcal{D}z_{i}=\frac{dz_{i}}{\sqrt{2\pi}}e^{-\frac{z_{i}^{2}}{2}}$.
Consequently, different replicas are decoupled and we have
\begin{align}
\int\prod_{a=1}^{n}d\boldsymbol{\Delta}^{a}e^{U}  = & \int\prod_{a=1}^{n}d\boldsymbol{\Delta}^{a}\exp\{-\lambda\beta\sum_{a}\left\Vert \boldsymbol{\triangle}^{a}\right\Vert ^{2}\nonumber \\
 & +\left(\hat{Q}-\frac{\hat{q}}{2}\right)\sum_{a}\left(\boldsymbol{\triangle}^{a}\right)^{T}\Lambda\boldsymbol{\triangle}^{a}\}\prod_{i}\int\mathcal{D}z_{i}e^{\sqrt{\gamma_{i}\hat{q}}z_{i}\left(\sum_{a}\triangle_{i}^{a}\right)}\nonumber \\
  = & \int\prod_{i}\mathcal{D}z_{i}\int\prod_{a=1}^{n}d\boldsymbol{\triangle}^{a}\prod_{a}\exp\{-\lambda\beta\sum_{i}\left(\triangle_{i}^{a}\right)^{2}\nonumber \\
 & +\left(\hat{Q}-\frac{\hat{q}}{2}\right)\sum_{i}\gamma_{i}\left(\triangle_{i}^{a}\right)^{2}+\sum_{i}\sqrt{\gamma_{i}\hat{q}}z_{i}\triangle_{i}^{a}\}\nonumber \\
  = & \int\prod_{i}\mathcal{D}z_{i}\int\prod_{a=1}^{n}d\boldsymbol{\triangle}^{a}\prod_{a}\exp\{\sum_{i}\left[\left(\hat{Q}-\frac{\hat{q}}{2}\right)\gamma_{i}-\lambda\beta\right]\left(\triangle_{i}^{a}\right)^{2}\nonumber \\
 & +\sum_{i}\left(\sqrt{\gamma_{i}\hat{q}}z_{i}\right)\triangle_{i}^{a}\}.\label{eq:eU-result1}
\end{align}
Since
\begin{equation}
\int dxe^{-Ax^{2}+Bx}=\sqrt{\frac{\pi}{A}}e^{\frac{B^{2}}{4A}},\label{eq:indentity-1}
\end{equation}
then
\begin{align}
 & \int d\triangle_{i}^{a}\exp\left\{ \left[\left(\hat{Q}-\frac{\hat{q}}{2}\right)\gamma_{i}-\lambda\beta\right]\left(\triangle_{i}^{a}\right)^{2}+\left(\sqrt{\gamma_{i}\hat{q}}z_{i}\right)\triangle_{i}^{a}\right\} \nonumber \\
= & \sqrt{\frac{2\pi}{\gamma_{i}\left(\hat{q}-2\hat{Q}+2\lambda\beta/\gamma_{i}\right)}}\exp\left[\frac{1}{2}\frac{\left(\sqrt{\hat{q}}z_{i}\right)^{2}}{\hat{q}-2\hat{Q}+2\lambda\beta/\gamma_{i}}\right].\label{eq:interg-wai}
\end{align}
Substituting (\ref{eq:interg-wai}) into (\ref{eq:eU-result1}), we
have
\begin{align}
\int\prod_{a=1}^{n}d\boldsymbol{\triangle}^{a}e^{U} & =\int\prod_{i}\mathcal{D}z_{i}\times\nonumber \\
 & \exp\left\{ n\left[\sum_{i}\frac{1}{2}\frac{\left(\sqrt{\hat{q}}z_{i}\right)^{2}}{\hat{q}-2\hat{Q}+2\lambda\beta/\gamma_{i}}+\frac{1}{2}\log2\pi-\frac{1}{2}\log\gamma_{i}-\frac{1}{2}\log\left(\hat{q}-2\hat{Q}+2\lambda\beta/\gamma_{i}\right)\right]\right\} .
\end{align}
Consequently, the original high-dimensional integration reduces to
a product of one-dimensional integrations w.r.t. $z_{i}$, independently.
\begin{align}
 & \int\mathcal{D}z_{i}\exp\left[\frac{1}{2}\frac{n\left(\sqrt{\hat{q}}z_{i}\right)^{2}}{\hat{q}-2\hat{Q}+2\lambda\beta/\gamma_{i}}\right]
= \sqrt{\frac{1}{1-\frac{n\hat{q}}{\hat{q}-2\hat{Q}+2\lambda\beta/\gamma_{i}}}}.
\end{align}
Then, we obtain 
\begin{align}
\int\prod_{a=1}^{n}d\boldsymbol{\triangle}^{a}e^{U} & =\exp\{-\frac{1}{2}\sum_{i}\log\left(1-\frac{n\hat{q}}{\hat{q}-2\hat{Q}+2\lambda\beta/\gamma_{i}}\right)\nonumber \\
 & +\frac{N}{2}\log2\pi-\frac{1}{2}\sum_{i}\log\gamma_{i}-\frac{1}{2}\sum_{i}\log\left(\hat{q}-2\hat{Q}+2\lambda\beta/\gamma_{i}\right)\}.
\end{align}
Thus, 
\begin{align}
\lim_{n\rightarrow0}\frac{S}{n} & =\underset{\hat{q},\hat{Q}}{\textrm{Extr}}\left\{ \lim_{n\rightarrow0}\frac{\log\int d\hat{Q}d\hat{q}e^{S_{X}}\int\prod_{a=1}^{n}d\boldsymbol{\triangle}^{a}e^{U}}{Nn}\right\} \nonumber \\
 & =\underset{\hat{q},\hat{Q}}{\textrm{Extr}}\{-\left(\lambda\beta\sum_{j\in\Psi_d}\bar{J}_{j}^{2}+\hat{Q}Q-\frac{1}{2}\hat{q}q\right)+\frac{\hat{q}}{2N}\sum_{i}\frac{1}{\hat{q}-2\hat{Q}+2\lambda\beta/\gamma_{i}}\nonumber \\
 & +\frac{1}{2}\log2\pi-\frac{1}{2N}\sum_{i}\log\gamma_{i}-\frac{1}{2N}\sum_{i}\log\left(\hat{q}-2\hat{Q}+2\lambda\beta/\gamma_{i}\right)\},\label{eq:eNS/n-result}
\end{align}
where $\underset{\hat{q},\hat{Q}}{\textrm{Extr}}\{\cdot\}$ denotes the
extreme operation over $\hat{q},\hat{Q}$. 
The summation in (\ref{eq:eNS/n-result}) is difficult to calculate. However, in the large system
limit, the summation converges to the integration, which
leads to
\begin{align}
\lim_{n\rightarrow0}\frac{S}{n}  = & \underset{\hat{q},\hat{Q}}{\textrm{Extr}}\{-\left(\lambda\beta\sum_{j\in\Psi_d}\bar{J}_{j}^{2}+\hat{Q}Q-\frac{1}{2}\hat{q}q\right)+\frac{\hat{q}}{2\beta}G_{1}\left(\frac{\hat{q}-2\hat{Q}}{\beta}\right)\nonumber \\
 & +\frac{1}{2}\log2\pi-\frac{1}{2N}\textrm{Tr}\log\left(\boldsymbol{C}^{\backslash0}\right)^{-1}-\frac{1}{2}G_{2}\left(\frac{\hat{q}-2\hat{Q}}{\beta}\right)-\frac{1}{2}\log\beta\},\label{eq:eNS/n-result-2}
\end{align}
where
\begin{align}
G_{1}\left(x\right)  = & \int\frac{d\eta\rho\left(\eta\right)}{x+2\lambda\eta},\label{eq:G1x-1}\\
G_{2}\left(x\right)  = & \int d\eta\rho\left(\eta\right)\log\left(x+2\lambda\eta\right).
\end{align}
Consequently, in (\ref{eq:eNS/n-result-2}), the extremization w.r.t. $\hat{q},\;\hat{Q}$ leads
to
\begin{equation}
\begin{cases}
q=-\frac{\hat{q}}{\beta^{2}}G_{1}^{'}\left(\frac{\hat{q}-2\hat{Q}}{\beta}\right),\\
Q-q=\frac{1}{\beta}G_{1}\left(\frac{\hat{q}-2\hat{Q}}{\beta}\right). 
\end{cases} \label{eq:fixed-point equation-1}
\end{equation}
Therefore, we obtain 
\begin{align}
\lim_{n\rightarrow0}\frac{S}{n}  = & -\lambda\beta\sum_{j\in\Psi_d}\bar{J}_{j}^{2}+\frac{Q\beta}{2}G_{1}^{-1}\left(\beta\left(Q-q\right)\right)-\frac{1}{2}G_{2}\left(G_{1}^{-1}\left(\beta\left(Q-q\right)\right)\right)\nonumber \\
 & +\frac{1}{2}\log\frac{2\pi}{\beta}-\frac{1}{2N}\textrm{Tr}\log\left(\boldsymbol{C}^{\backslash0}\right)^{-1}.
\end{align}

%%%%%%%%%%%%%%%%%%%%%%%%%%%%%%%%%%%%%%%%%%%%%%%%%%%%%
%%%%%%%%%%%%%%%%%%%%%%%%%%%%%%%%%%%%%%%%%%%%%%%%%%%%%
\subsection{\label{subsec:appendix-L}Computation of $L$}
The definition of $L$ is given in (\ref{eq:L-part-df-1}), which
is 
\begin{equation}
L\doteq\sum_{s_{0},\boldsymbol{s}_{\Psi_d}}P\left(s_{0},\boldsymbol{s}_{\Psi_d}|J^{*}\right)\int\prod_{a=1}^{n}dh_{\triangle}^{a}P_{\textrm{cav}}\left(\left\{ h_{\triangle}^{a}\right\} _{a}|\left\{ \boldsymbol{\Delta}^{a}\right\} _{a}\right)e^{-\beta\sum_{a=1}^{n}\Phi\left(s_{0}\left(\sum_{j\in\Psi_d}\text{\ensuremath{\bar{J}_{j}}}s_{j}+h_{\triangle}^{a}\right)\right)}.\label{eq:L-def-appendix}
\end{equation}
Using the cavity method, the local fields $h_{\triangle}^{a}, a=1,\ldots,n$
follow a joint Gaussian distribution with zero mean (paramagnetic
assumption) and covariances as
\begin{equation}
\left\langle h_{a}h_{b}\right\rangle ^{\backslash0}=Q\delta_{ab}+\left(1-\delta_{ab}\right)q. \label{eq:cov-h-def}
\end{equation}
Then, we can introduce two auxiliary i.i.d. Gaussian random variables
$v_{a},z$ with zero mean and unit variance, by which the local fields
can be written in a compact form
\begin{equation}
h_{a}=\sqrt{Q-q}v_{a}+\sqrt{q}z\label{eq:local-field-compactform}
\end{equation}
so that $L$ in (\ref{eq:L-def-appendix}) can be equivalently written
as
\begin{align}
L & \doteq\sum_{s_{0},\boldsymbol{s}_{\Psi_d}}P\left(s_{0},\boldsymbol{s}_{\Psi_d}|J^{*}\right)\int\prod_{a=1}^{n}dh_{\triangle}^{a}P_{\textrm{cav}}\left(\left\{ h_{\triangle}^{a}\right\} _{a}|\left\{ \boldsymbol{\Delta}^{a}\right\} _{a}\right)e^{-\beta\sum_{a=1}^{n}\Phi\left(s_{0}\left(\sum_{j\in\Psi_d}\text{\ensuremath{\bar{J}_{j}}}s_{j}+h_{\triangle}^{a}\right)\right)}\nonumber \\
 & =\sum_{s_{0},\boldsymbol{s}_{\Psi_d}}P\left(s_{0},\boldsymbol{s}_{\Psi_d}|J^{*}\right)\int\mathcal{D}z\prod_{a}\mathcal{D}v_{a}e^{-\beta\sum_{a=1}^{n}\Phi\left(s_{0}\left(\sum_{j\in\Psi_d}\text{\ensuremath{\bar{J}_{j}}}s_{j}+\sqrt{Q-q}v_{a}+\sqrt{q}z\right)\right)}\nonumber \\
 & =\sum_{s_{0},\boldsymbol{s}_{\Psi_d}}P\left(s_{0},\boldsymbol{s}_{\Psi_d}|J^{*}\right)\int\mathcal{D}z\left[\underset{A}{\underbrace{\int\mathcal{D}ve^{-\beta\Phi\left(s_{0}\left(\sum_{j\in\Psi_d}\text{\ensuremath{\bar{J}_{j}}}s_{j}+\sqrt{Q-q}v+\sqrt{q}z\right)\right)}}}\right]^{n}\nonumber \\
 & =\sum_{s_{0},\boldsymbol{s}_{\Psi_d}}P\left(s_{0},\boldsymbol{s}_{\Psi_d}|J^{*}\right)E_{z}\left(A^{n}\right),
\end{align}
where $E_{z}\left(A^{n}\right)=\int\mathcal{D}z A^{n}$. Then, using the replica formula, we have
\begin{align}
\underset{n\rightarrow0}{\lim}\frac{1}{n}\log L & =\underset{n\rightarrow0}{\lim}\frac{\log\sum_{s_{0},\boldsymbol{s}_{\Psi_d}}P\left(s_{0},\boldsymbol{s}_{\Psi_d}|J^{*}\right)E_{z}\left(A^{n}\right)}{n}\nonumber \\
 & =E_{z}\left[\sum_{s_{0},\boldsymbol{s}_{\Psi_d}}P\left(s_{0},\boldsymbol{s}_{\Psi_d}|J^{*}\right)A\right]\nonumber \\
 & =\sum_{s_{0},\boldsymbol{s}_{\Psi_d}}P\left(s_{0},\boldsymbol{s}_{\Psi_d}|J^{*}\right)\int\mathcal{D}z\log\int\mathcal{D}ve^{-\beta\Phi\left(s_{0}\left(\sum_{j\in\Psi_d}\text{\ensuremath{\bar{J}_{j}}}s_{j}+\sqrt{Q-q}v+\sqrt{q}z\right)\right)}.\label{eq:logL/n-result}
\end{align}
To further simplify the result, let $y=s_{0}\left(\sum_{j\in\Psi_d}\text{\ensuremath{\bar{J}_{j}}}s_{j}+\sqrt{Q-q}v+\sqrt{q}z\right)$. Consequently, we obtain 
\begin{align}
 & \int\mathcal{D}ve^{-\beta\Phi\left(s_{0}\left(\sum_{j\in\Omega}\text{\ensuremath{\bar{J}_{j}}}s_{j}+\sqrt{Q-q}v+\sqrt{q}z\right)\right)}\nonumber \\
= & \int\frac{dv}{\sqrt{2\pi}}e^{-\frac{v^{2}}{2}}e^{-\beta\Phi\left(y\right)}\nonumber \\
= & \int\frac{dy}{\sqrt{2\pi\left(Q-q\right)}}e^{-\frac{\left[y-s_{0}\left(\sum_{j\in\Omega}\text{\ensuremath{\bar{J}_{j}}}s_{j}+\sqrt{q}z\right)\right]^{2}}{2\left(Q-q\right)}}e^{-\beta\Phi\left(y\right)},\label{eq:temp-results1}
\end{align}
so that
\begin{align}
\underset{n\rightarrow0}{\lim}\frac{1}{n}\log L & =\sum_{s_{0},\boldsymbol{s}_{\Psi_d}}P\left(s_{0},\boldsymbol{s}_{\Psi_d}|J^{*}\right)\int\mathcal{D}z\log\int\frac{dy}{\sqrt{2\pi\left(Q-q\right)}}e^{-\frac{\left[y-s_{0}\left(\sum_{j\in\Psi_d}\text{\ensuremath{\bar{J}_{j}}}s_{j}+\sqrt{q}z\right)\right]^{2}}{2\left(Q-q\right)}}e^{-\beta\Phi\left(y\right)}\nonumber \\
 & =\sum_{s_{0},\boldsymbol{s}_{\Psi_d}}P\left(s_{0},\boldsymbol{s}_{\Psi_d}|J^{*}\right)\int\mathcal{D}z\underset{y}{\max}\left[-\frac{\left(y-s_{0}\left(\sqrt{q}z+\sum_{j\in\Psi_d}\text{\ensuremath{\bar{J}_{j}}}s_{j}\right)\right)^{2}}{2\left(Q-q\right)}-\beta\Phi\left(y\right)\right].
\end{align}

%%%%%%%%%%%%%%%%%%%%%%%%%%%%%%%%%%%%%%%%%%%%%%%%%%%%%
%%%%%%%%%%%%%%%%%%%%%%%%%%%%%%%%%%%%%%%%%%%%%%%%%%%%%
\subsection{Derivation of Macroscopic Parameters}
\label{Appendix-MacroParameters-auxiliarymethod}
We use the technique of auxiliary variables in \cite{bachschmid2017statistical} by introducing the term
$h_{R}\sum_{a=1}^{n}\left\Vert \boldsymbol{W}^{a}\right\Vert ^{2}$
into $\left[Z^{n}\right]_{\mathcal{D}^{M}}$. Then, following the same procedure
as that in \ref{subsec:Appendix-S}, we obtain 
\begin{align}
\lim_{n\rightarrow0}\frac{S}{n} & =\underset{\hat{q},\hat{Q}}{\textrm{Extr}}\{-\lambda\beta\sum_{j\in\Psi_d}\bar{J}_{j}^{2}-\hat{Q}Q+\frac{1}{2}\hat{q}q+\frac{1}{2}\frac{1}{N}\sum_{i}\frac{\gamma_{i}\hat{q}}{\left(\hat{q}-2\hat{Q}\right)\gamma_{i}+2\left(\lambda\beta-h_{R}\right)}\nonumber \\
 & -\frac{1}{2N}\sum_{i}\log\left(\hat{q}-2\hat{Q}+2\left(\lambda\beta-h_{R}\right)/\gamma_{i}\right)-\frac{1}{2N}\sum_{i}\log\gamma_{i}\}.
\end{align}
Thus, the macroscopic parameter $R=\frac{1}{N}\sum_{j\in\bar{\Psi}_d}\triangle_{j}^{2}$
can be derived using the derivative of the free energy, i.e., 
\begin{align}
R & =\lim_{h_{R}\rightarrow0}\frac{\partial}{\partial h_{R}}\lim_{n\rightarrow0}\frac{S}{n}\nonumber \\
 & =\frac{1}{N}\sum_{i}\frac{\hat{q}/\gamma_{i}}{\left(\hat{q}-2\hat{Q}+2\lambda\beta/\gamma_{i}\right)^{2}}+\frac{1}{N}\sum\frac{1/\gamma_{i}}{\hat{q}-2\hat{Q}+2\lambda\beta/\gamma_{i}}\nonumber \\
 & =\frac{\hat{q}}{\beta^{2}}\frac{1}{N}\sum_{i}\frac{1/\gamma_{i}}{\left(\frac{\hat{q}-2\hat{Q}}{\beta}+2\lambda/\gamma_{i}\right)^{2}}+\frac{1}{\beta}\frac{1}{N}\sum\frac{1/\gamma_{i}}{\frac{\hat{q}-2\hat{Q}}{\beta}+2\lambda/\gamma_{i}}\nonumber \\
 & =\frac{\hat{q}}{\beta^{2}}\frac{1}{N}\sum_{i}\frac{1/\gamma_{i}}{\left(\frac{\hat{q}-2\hat{Q}}{\beta}+2\lambda/\gamma_{i}\right)^{2}}+\underset{\rightarrow0\;\left(\beta\rightarrow\infty\right)}{\underbrace{\frac{1}{\beta}\frac{1}{N}\sum\frac{1/\gamma_{i}}{a+2\lambda/\gamma_{i}}}}\nonumber \\
 & =q\frac{G_{3}^{'}\left(\kappa\right)}{G_{1}^{'}\left(\kappa\right)},\label{eq:R-results}
\end{align}
where
\begin{equation}
G_{3}\left(x\right)=\int\frac{d\eta\rho\left(\eta\right)\eta}{\left(x+2\lambda\eta\right)}.\label{eq:G3-x-1}
\end{equation}

%%%%%%%%%%%%%%%%%%%%%%%%%%%%%%%%%%%%%%%%%%%%%%%%%%%%%
%%%%%%%%%%%%%%%%%%%%%%%%%%%%%%%%%%%%%%%%%%%%%%%%%%%%%
\subsection{{\label{subsec:Numerical-Solutions}Numerical Solutions}}
In general, there is no analytic solution to the EOS equations in Section \ref{EOS-subsec}, but they can be easily solved using numerical methods. 

First, we compute $\hat{y}$ by substituting $\Phi\left(y\right)=\left(y-1\right)^{2}$
into (\ref{eq:y-hat-def}), which yields
\begin{equation}
\hat{y}=\frac{s_{0}\left(\sqrt{Q}z+\sum_{j\in\Psi_d}\text{\ensuremath{\bar{J}_{j}}}s_{j}\right)+2G_{1}\left(\kappa\right)}{1+2G_{1}\left(\kappa\right)},\label{eq:y-hat-result}
\end{equation}
from which we obtain
\begin{align}
\begin{cases}
\int\mathcal{D}z\frac{\partial\Phi\left(y\right)}{\partial y}\mid_{y=\hat{y}}=2\frac{s_{0}\sum_{j\in\Psi_d}\text{\ensuremath{\bar{J}_{j}}}s_{j}-1}{1+2G_{1}\left(\kappa\right)},\\
\int\mathcal{D}zz\frac{\partial\Phi\left(y\right)}{\partial y}\mid_{y=\hat{y}}=\frac{2s_{0}\sqrt{Q}}{1+2G_{1}\left(\kappa\right)},\\
\int\mathcal{D}z\left(\frac{\partial\Phi\left(y\right)}{\partial y}\mid_{y=\hat{y}}\right)^{2}=4\frac{Q+\left(\sum_{j\in\Psi_d}\text{\ensuremath{\bar{J}_{j}}}s_{j}\right)^{2}-2s_{0}\left(\sum_{j\in\Psi_d}\text{\ensuremath{\bar{J}_{j}}}s_{j}\right)+1}{\left[1+2G_{1}\left(\kappa\right)\right]^{2}}.
\end{cases}
\end{align}
The mean estimates $\text{\ensuremath{\left(\bar{J}_{j}\right) } }_{j\in\Psi_d}$
can also be evaluated by the extremization condition, i.e.,
\begin{align}
0 & =2\lambda\bar{J}_{j}+\alpha\sum_{s_{0},\boldsymbol{s}_{\Psi_d}}P\left(s_{0},\boldsymbol{s}_{\Psi_d}|J^{*}\right)\int\mathcal{D}z\frac{\partial\Phi\left(y\right)}{\partial y}\mid_{y=\hat{y}}s_{0}s_{j},\;j\in\Psi_d.\label{eq:J-bar-condition-1}
\end{align}
Hence, the mean estimates $\text{\ensuremath{\left(\bar{J}_{j}\right) } }_{j\in\Psi_d}$
can be evaluated from (\ref{eq:J-bar-condition-1}) by solving the
linear equations
\begin{align}
\left(1+2\lambda/\kappa\right)\bar{J}_{j}+\sum_{i\in\Psi_d,i\neq j}\bar{J}_{i}\left\langle s_{i}s_{j}\right\rangle -\left\langle s_{0}s_{j}\right\rangle =0,\; {j\in\Psi_d},\label{eq:J-mean-linear-equation}
\end{align}
where $\left\langle s_{i}s_{j}\right\rangle $ denotes the average w.r.t.
the joint distribution $P\left(s_{0},\boldsymbol{s}_{\Psi_d}|J^{*}\right)$. 

The macroscopic parameters $\kappa$ and $Q$ can be obtained by numerically solving
the following equations
\begin{align}
0 = & \kappa-\frac{2\alpha}{1+2G_{1}\left(\kappa\right)},\\
Q= & \frac{4\alpha G_{1}^{'}\left(\kappa\right)\left(2\left\langle s_{0}\sum_{j\in\Psi_d}\text{\ensuremath{\bar{J}_{j}}}s_{j}\right\rangle -\left\langle \left(\sum_{j\in\Psi_d}\text{\ensuremath{\bar{J}_{j}}}s_{j}\right)^{2}\right\rangle -1\right)}{\left(1+2G_{1}\left(\kappa\right)\right)^{2}+4\alpha G_{1}^{'}\left(\kappa\right)}.
\label{eq:q and a equation}
\end{align}

% Note that to obtain $\hat{b}$, we have

% \begin{equation}
% \hat{b}=\frac{\kappa\left\langle s_{0}\sum_{j\in\Omega_{1}}\text{\ensuremath{J_{j}^{*}}}s_{j}\right\rangle -\kappa\left\langle \left(\sum_{j\in\Omega_{1}}\text{\ensuremath{J_{j}^{*}}}s_{j}\right)\left(\sum_{i\in\Psi_d\setminus\Omega_{1}}\text{\ensuremath{\bar{J}_{i}}}s_{i}\right)\right\rangle }{\kappa\left\langle \left(s_{0}\sum_{j\in\Omega_{1}}\text{\ensuremath{J_{j}^{*}}}s_{j}\right)^{2}\right\rangle +2\lambda cK^{2}}.\label{eq:bias-estimate}
% \end{equation}

%%%%%%%%%%%%%%%%%%%%%%%%%%%%%%%%%%%%%%%%%%%%%%%%%%%%%
%%%%%%%%%%%%%%%%%%%%%%%%%%%%%%%%%%%%%%%%%%%%%%%%%%%%%
\subsection{\label{subsec:Eigenvalue-distribution}Eigenvalue Distribution}
From the replica analysis presented, the learning performance will depend
on the eigenvalue distribution (EVD) $\rho\left(\eta\right)$ of the
inverse correlation matrix $\left(\boldsymbol{C}^{\backslash0}\right)^{-1}$.
In general, it is difficult to obtain this EVD; however, for a particular
teacher network, we can obtain the analytic solution of $\rho\left(\eta\right)$.
In this section, we illustrate how to compute $\rho\left(\eta\right)$
of the random regular (RR) graph as a representative example of
sparse tree-like graphs. Assume that as $N\rightarrow\infty$, the EVD
of $\left(\boldsymbol{C}^{\backslash0}\right)^{-1}$ approaches
that of $\boldsymbol{C}^{-1}$ in the large system
limit, where $\boldsymbol{C}$ is the correlation matrix that corresponds
to the teacher spin system.
The Gibbs free energy is defined as
\begin{equation}
G\left(\boldsymbol{m}\right)=\underset{\boldsymbol{\theta}}{\max}\left\{ \boldsymbol{\theta}^{T}\boldsymbol{m}-\log Z\left(\boldsymbol{\theta}\right)\right\} ,\label{eq:Gibbs-Gm-1}
\end{equation}
where $Z\left(\boldsymbol{\theta}\right)=\sum_{\boldsymbol{s}}e^{\sum_{i<j}J_{ij}s_{i}s_{j}+\sum_{i}\theta_{i}s_{i}}$.
It can be verified that the Hessian of $G\left(\boldsymbol{m}\right)$
is equal to the inverse correlation matrix, i.e., $\left[\boldsymbol{C}^{-1}\right]_{ij}=\frac{\partial G\left(\boldsymbol{m}\right)}{\partial m_{i}\partial m_{j}}$.
Consequently, we can focus on the computation of $G\left(\boldsymbol{m}\right)$
to obtain the EVD of $\boldsymbol{C}^{-1}$ . The RR
graph is characterized by a connectivity parameter $c$ and constant
coupling strength $K$. The inverse correlation matrix can be computed
from the Hessian of the Gibbs free energy \cite{Abbara2019c, ricci2012bethe, nguyen2012bethe} as 
\begin{align}
\left[\boldsymbol{C}^{-1}\right]_{ij} & =\frac{\partial G\left(\boldsymbol{m}\right)}{\partial m_{i}\partial m_{j}}\nonumber \\
 & =\left(\frac{c}{1-\tanh^{2}K}-c+1\right)\delta_{ij}-\frac{\tanh\left(J_{ij}\right)}{1-\tanh^{2}\left(J_{ij}\right)}\left(1-\delta_{ij}\right),\label{eq:ivnerse-correlation-mat-result}
\end{align}
and in matrix form, we have
\begin{equation}
\boldsymbol{C}^{-1}=\left(\frac{c}{1-\tanh^{2}K}-c+1\right)\mathbf{I}-\frac{\tanh\left(\boldsymbol{J}\right)}{1-\tanh^{2}\left(\boldsymbol{J}\right)}.\label{eq:inverse-corre-mat-C-1}
\end{equation}
Since the matrix $\frac{\tanh\left(\boldsymbol{J}\right)}{1-\tanh^{2}\left(\boldsymbol{J}\right)}$
is also a sparse coupling matrix with constant coupling strength $K_{1}=\frac{\tanh\left(K\right)}{1-\tanh^{2}\left(K\right)}$
and fixed connectivity $c$, the corresponding
eigenvalue (denoted as $\xi$) distribution can be calculated as \cite{mckay1981expected}
\begin{equation}
\rho_{\xi}\left(\xi\right)=\frac{c\sqrt{4K_{1}^{2}\left(c-1\right)-\xi^{2}}}{2\pi\left(K_{1}^{2}c^{2}-\xi^{2}\right)},\;\left|\xi\right|\leq2K_{1}\sqrt{c-1}. \label{eq:EVD-kesi}
\end{equation}
From (\ref{eq:inverse-corre-mat-C-1}), the eigenvalue $\eta$
of $\boldsymbol{C}^{-1}$ is
\begin{equation}
\eta_{i}=\frac{c}{1-\tanh^{2}K}-c+1-\xi_{i},\label{eq:eigen-value-result}
\end{equation}
which, when combined with (\ref{eq:EVD-kesi}), 
readily yields the EVD of $\eta$ as $N\rightarrow\infty$ as follows:
\begin{align}
\rho\left(\eta\right) & =\rho_{\xi}\left(\frac{c}{1-\tanh^{2}K}-c+1-\eta\right)\nonumber \\
 & =\frac{c\sqrt{4\left(\frac{\tanh\left(K\right)}{1-\tanh^{2}\left(K\right)}\right)^{2}\left(c-1\right)-\left(\frac{c}{1-\tanh^{2}K}-c+1-\eta\right)^{2}}}{2\pi\left(\left(\frac{\tanh\left(K\right)}{1-\tanh^{2}\left(K\right)}\right)^{2}c^{2}-\left(\frac{c}{1-\tanh^{2}K}-c+1-\eta\right)^{2}\right)},\label{eq:EVD-nita-results}
\end{align}
where $\eta\in\left[\frac{c}{1-\tanh^{2}K}-c+1-\frac{2\tanh\left(K\right)\sqrt{c-1}}{1-\tanh^{2}\left(K\right)},\frac{c}{1-\tanh^{2}K}-c+1+\frac{2\tanh\left(K\right)\sqrt{c-1}}{1-\tanh^{2}\left(K\right)}\right]$.

%%%%%%%%%%%%%%%%%%%%%%%%%%%%%%%%%%%%%%%%%%%%%%%%%%%%%
%%%%%%%%%%%%%%%%%%%%%%%%%%%%%%%%%%%%%%%%%%%%%%%%%%%%%
\subsection{\label{subsec:Lemma1-proof}Proof of Theorem 1}
%%\begin{proof}
Although the proof in the previous study (the one in Sec. 3.3 in \cite{Abbara2019c}) can be applied to the present case, we provide another proof by employing some specific properties of the linear regression, because some steps in this proof are essential for Theorem 2, which is beyond the applicable range of the proof in \cite{Abbara2019c}. Note that the advantage of the proof in \cite{Abbara2019c}  is its generality: an arbitrary cost function and the nonzero external fields are treated.

Specifically, in this case, the linear equations in (\ref{eq:J-mean-linear-equation-1})
reduce to 
\begin{align}
\sum_{i\in\Psi_d}\bar{J}_{i}\left\langle s_{i}s_{j}\right\rangle =\left\langle s_{0}s_{j}\right\rangle , \; j\in\Psi_d.\label{eq:J-mean-linear-equation-lamda0}
\end{align}
In matrix form, 
\begin{align}
\boldsymbol{C}_d\boldsymbol{\bar{J}} & =\boldsymbol{b},\label{eq:Linear-equation-1-1}\\
\boldsymbol{\bar{J}}=\left[\begin{array}{c}
\bar{J}_{1}\\
\bar{J}_{2}\\
\vdots\\
\bar{J}_{\left|\Psi_d\right|}
\end{array}\right], \; & \boldsymbol{b}=\left[\begin{array}{c}
\left\langle s_{0}s_{1}\right\rangle \\
\left\langle s_{0}s_{2}\right\rangle \\
\vdots\\
\left\langle s_{0}s_{\left|\Psi_d\right|}\right\rangle 
\end{array}\right],
\end{align}
where $\boldsymbol{C}_d=\left\{ \left\langle s_{i}s_{j}\right\rangle \right\} _{i,j\in\Psi_d}$
is the correlation matrix of spins $s_{j},j\in\Psi_d$. Consequently, the estimates
$\boldsymbol{\bar{J}}=\text{\ensuremath{\left(  \bar{J}_{j}\right) } }_{j\in\Psi_d}$ can
be computed as $\boldsymbol{\bar{J}}=\boldsymbol{C}_d^{-1}\boldsymbol{b}$. 

On the one hand, the full correlation matrix $\boldsymbol{C}_{0d}=\left\{ \left\langle s_{i}s_{j}\right\rangle \right\} _{i,j\in\left\{ 0,\Psi_d\right\} }$ of spins $s_j, \;j \in\left\{ 0,\Psi_d\right\}$
can be represented as
\begin{equation}
\boldsymbol{C}_{0d}=\left[\begin{array}{cc}
1 & \boldsymbol{b}^{T}\\
\boldsymbol{b} & \boldsymbol{C}_d
\end{array}\right].\label{eq:Covariance-matrix}
\end{equation}
Thus, according to the block matrix inversion lemma, the inverse correlation matrix can be computed as
\begin{align}
\boldsymbol{C}_{0d}^{-1} & =\left[\begin{array}{cc}
F_{11}^{-1} & -F_{11}^{-1}\boldsymbol{\bar{J}}^{T}\\
-\boldsymbol{\bar{J}}F_{11}^{-1} & F_{22}^{-1}
\end{array}\right],\label{eq:Covariance-matrix-1}
\end{align}
where $F_{11}=1-\boldsymbol{b}^{T}\boldsymbol{\bar{J}},F_{22}=\boldsymbol{C}_d-\boldsymbol{b}\boldsymbol{b}^{T}.$ 

On the other hand, for a sparse tree graph where $s_{i}$ has connectivity
$c_{i}$ with true coupling $J_{ij}$ with spin $s_{j}$, the inverse
correlation matrix can be computed from the Hessian of the Gibbs free
energy as \cite{Abbara2019c, ricci2012bethe, nguyen2012bethe}
\begin{align}
\left[\boldsymbol{C}_{0d}^{-1}\right]_{ij} & =\left(\sum_{k\in\partial i}\frac{1}{1-\tanh^{2}J_{ik}}-c_{i}+1\right)\delta_{ij}-\frac{\tanh\left(J_{ij}\right)}{1-\tanh^{2}J_{ij}}\left(1-\delta_{ij}\right).\label{eq:C0-inverse-matrix-1-v2}
\end{align}
The two representations of $\boldsymbol{C}_{0d}^{-1}$ in (\ref{eq:Covariance-matrix-1})
and (\ref{eq:C0-inverse-matrix-1-v2}) are equivalent; hence, the
corresponding elements should be equal to each other. Specifically, we
are interested in the first row, which corresponds to spin $s_{0}$.
Denote $J_{0j}=J_{j}^{*}$ as the true couplings associated with spin
$s_{0}$ in the teacher network. Then, by the definition of $\Omega_{1}$,
we have $J_{j}^{*}=0,j\notin\Omega_{1}$. Assuming that $c_{0}=c=\left|\Omega_{1}\right|$,
by comparing (\ref{eq:Covariance-matrix-1})
and (\ref{eq:C0-inverse-matrix-1-v2}), it is easy to obtain
\begin{align}
F_{11}^{-1}=&\sum_{k\in\Omega_{1}}\frac{1}{1-\tanh^{2}J_{k}^{*}}-c+1,\\
\bar{J}_{j}F_{11}^{-1}=&\frac{\tanh\left(J_{j}^{*}\right)}{1-\tanh\left(J_{j}^{*}\right)},
\label{eq:Jbar-condition}
\end{align}
which is the same as (\ref{eq:solution-Jbar-lamda=00003D0-1}). The
result of (\ref{eq:solution-Jbar-lamda=00003D0}) can be readily
obtained for constant couplings by substituting $J_{j}^{*}=K\textrm{sign}(J_{j}^{*}),\;j\in\Omega_{1}$,
which completes the proof. 
%%\end{proof}

%%%%%%%%%%%%%%%%%%%%%%%%%%%%%%%%%%%%%%%%%%%%%%%%%%%%%
%%%%%%%%%%%%%%%%%%%%%%%%%%%%%%%%%%%%%%%%%%%%%%%%%%%%%
\subsection{\label{subsec:Lemma2-proof}Proof of Theorem 2}
%% \begin{proof}
In this case, the estimate of $\boldsymbol{\bar{J}}=\text{\ensuremath{\left( \bar{J}_{j}\right) } }_{j\in\Psi_d}$
is the solution to the following linear equations:
\begin{align}
\left(\boldsymbol{C}_d+\frac{2\lambda}{\kappa}\boldsymbol{I}\right)\boldsymbol{\bar{J}} & =\boldsymbol{b},\label{eq:Linear-equation-regularize}
\end{align}
where $\boldsymbol{I}$ is the identity matrix. To evaluate the decay speed of $\text{\ensuremath{\left(  \bar{J}_{j}\right) } }_{j\in\Psi_d}$
with the distance from $s_{0}$, we can compute  the decay speed
of $\text{\ensuremath{\left(  \bar{J}_{j}\right) } }_{j\in\Psi_d}$
for two general NN spins $s_{i}$ and $s_{j}$ with
distance 1. For notational simplicity, denote $\textrm{Dist}\left(s_{i},s_{j}\right)$
as the distance between two spins $s_{i}$ and $s_{j}$ in the teacher
Ising system. Then, for two NN spins $s_{i}$ and $s_{j}$,
$\textrm{Dist}\left(s_{i},s_{j}\right)=1$. Without loss of generality, using the gauge symmetry, we can assume that all the true couplings of the teacher Ising spin
system are non-negative when the external field is absent and the paramagnet assumption holds; hence, each element in $\boldsymbol{C}_d$
and $\boldsymbol{b}$ is positive. Assuming that $\textrm{Dist}\left(s_{i},s_{0}\right)=d-1$
and $\textrm{Dist}\left(s_{j},s_{0}\right)=d$, then $\left\langle s_{0}s_{i}\right\rangle =\theta^{d-1}$
and $\left\langle s_{0}s_{j}\right\rangle =\theta^{d}$. In general,
there are two cases. 

\textbf{Case 1}: $s_{j}$ is a leaf spin, which means that $s_{j}$
is only directly connected to its parent spin $s_{i}$ and has no
children spins. 

In this case, for any other spin $s_{k},k\in\Psi_d,k\neq j$ , we have
$\textrm{Dist}\left(s_{j},s_{k}\right)=\textrm{Dist}\left(s_{i},s_{k}\right)+1$.
Then, the associated rows corresponding to $s_{i}$ and $s_{j}$ in
(\ref{eq:Linear-equation-regularize}) can be written as follows:
\begin{equation}
\begin{cases}
\left(1+\frac{2\lambda}{\kappa}\right)\bar{J}_{i}+\theta\bar{J}_{j}+\sum_{k\in\Psi_d,k\neq i,j}\theta^{d_{ki}}\bar{J}_{k}=\theta^{d-1},\\
\theta\bar{J}_{i}+\left(1+\frac{2\lambda}{\kappa}\right)\bar{J}_{j}+\sum_{k\in\Psi_d,k\neq i,j}\theta^{d_{ki}+1}\bar{J}_{k}=\theta^{d},
\end{cases},\label{eq:Linear-equation-mat-Leafnode}
\end{equation}
where $d_{ki}=\textrm{Dist}\left(s_{i},s_{k}\right)$ and $d_{kj}=\textrm{Dist}\left(s_{j},s_{k}\right)$.
From (\ref{eq:Linear-equation-mat-Leafnode}),
we can easily obtain 

\begin{equation}
0<\frac{\bar{J}_{j}}{\bar{J}_{i}}=\theta\frac{\left(D-1\right)}{D-\theta^{2}}<\theta,\label{eq:decay-ratio-Leaf}
\end{equation}
where $D=1+\frac{2\lambda}{\kappa}>1$, which implies that the ratio
of the magnitude of $\bar{J}_{j}$ to that of its parent node $\bar{J}_{i}$ is
smaller than $\theta$ whenever $s_{j}$ is a leaf spin. Meanwhile,
the signs of $\bar{J}_{j}$ and $\bar{J}_{i}$ are always the same, i.e.,
$s_{i}$ has the same sign as its children spin $s_{j}$ when $s_{j}$
is a leaf spin. 

\textbf{Case }2: $s_{j}$ is not a leaf spin but has its own direct
children spins.

Denote $\Phi$ as the set of children spins of $s_{j}$. Then, for
any spin $s_{m},m\in\Phi$ , we have $\textrm{Dist}\left(s_{j},s_{m}\right)=\textrm{Dist}\left(s_{i},s_{m}\right)-1$.
For any other spin $s_{k},k\in\Psi_d/\Phi,k\neq j$ , we have
$\textrm{Dist}\left(s_{j},s_{k}\right)=\textrm{Dist}\left(s_{i},s_{k}\right)+1$.
Consequently, the associated rows corresponding to $s_{i}$ and $s_{j}$
in (\ref{eq:Linear-equation-regularize}) can be written as follows:
\begin{equation}
\begin{cases}
\left(1+\frac{2\lambda}{\kappa}\right)\bar{J}_{i}+\theta\bar{J}_{j}+\sum_{k\in\Psi_d/\Phi,k\neq i,j}\theta^{d_{ki}}\bar{J}_{k}+\sum_{m\in\Phi}\theta^{d_{mj}+1}\bar{J}_{m}=\theta^{d-1},\\
\theta\bar{J}_{i}+\left(1+\frac{2\lambda}{\kappa}\right)\bar{J}_{j}+\sum_{k\in\Psi_d/\Phi,k\neq i,j}\theta^{d_{ki}+1}\bar{J}_{k}+\sum_{m\in\Phi}\theta^{d_{mj}}\bar{J}_{m}=\theta^{d}.
\end{cases}\label{eq:Linear-equation-mat-general-node}
\end{equation}
Then, denoting $D=1+\frac{2\lambda}{\kappa}>1$, we obtain
\begin{align}
\frac{\bar{J}_{i}}{\bar{J}_{j}} & =\frac{D-\theta^{2}}{\theta\left(D-1\right)}+\frac{1}{\bar{J}_{j}}\sum_{m\in\Phi}\bar{J}_{m}\left(\theta^{d_{mj}}-\theta^{d_{mj}+2}\right).\label{eq:decay-ratio-general-1}
\end{align}
Since $\Phi$ is the set of children spins of $s_{j}$, then for any
$\bar{J}_{m},m\in\Phi$, it can be deduced by induction that $\bar{J}_{m}$
has the same sign as $\bar{J}_{j}$ as follows. 

First, if $s_{m},m\in\Phi$ are all leaf spins, then using the result
of case 1, $\bar{J}_{m}$ has the same sign as $\bar{J}_{j}$, i.e.,
$\bar{J}_{m}/\bar{J}_{j}>0$. Second, if $s_{m},m\in\Phi$ is not
a leaf spin itself but has a leaf spin $s_{n}$, then $\bar{J}_{n}$
has the same sign as $\bar{J}_{m}$, and as with (\ref{eq:decay-ratio-general-1}),
we can obtain $\bar{J}_{j}/\bar{J}_{m}=\frac{D-\theta^{2}}{\theta\left(D-1\right)}+\frac{1}{\bar{J}_{m}}\bar{J}_{n}\left(\theta-\theta^{2}\right)>0$;
hence, $\bar{J}_{m}$ has the same sign as $\bar{J}_{j}$. By induction,
the estimates of the children spins $\bar{J}_{m},m\in\Phi$ will all have
the same sign as $\bar{J}_{j}$. 

Consequently, since (\ref{eq:decay-ratio-general-1}), $\bar{J}_{m}/\bar{J}_{j}>0,m\in\Phi$,
and $0<\theta<1$, we have
\begin{align}
\frac{\bar{J}_{i}}{\bar{J}_{j}}> & \frac{D-\theta^{2}}{\theta\left(D-1\right)}>0\iff0<\frac{\bar{J}_{j}}{\bar{J}_{i}}<\frac{\theta\left(D-1\right)}{D-\theta^{2}},\label{eq:decay-ratio-generalnode}
\end{align}
which implies that the ratio of the magnitude of $\bar{J}_{j}$ to that of its
parent spin $\bar{J}_{i}$ is smaller than $\frac{\theta\left(D-1\right)}{D-\theta^{2}}$ for general $s_{j}$
when it is not a leaf spin. 

Summarizing both case 1 and case 2, for any two spins $s_{i}$ and
$s_{j}$ with distance $\textrm{Dist}\left(s_{i},s_{j}\right)=1$,
$\left|\frac{\bar{J}_{j}}{\bar{J}_{i}}\right|\leq \frac{\theta\left(D-1\right)}{D-\theta^{2}}$ always holds,
which completes the proof. 
%%\end{proof}

%%%%%%%%%%%%%%%%%%%%%%%%%%%%%%%%%%%%%%%%%%%%%%%%%%%%%
%%%%%%%%%%%%%%%%%%%%%%%%%%%%%%%%%%%%%%%%%%%%%%%%%%%%%
%%%%%%%%%%%%%%%%%%%%%%%%%%%%%%%%%%%%%%%%%%%%%%%%%%%%%
\bibliographystyle{unsrt}
\bibliography{main}

\end{document}